\documentclass[useAMS,usenatbib]{mn2e}
\usepackage{graphics}
\newcommand{\figdir}{./}
\newcommand{\ddd}{\,\mathrm{d}}
\title
[Final Analysis of ELAIS 15 $\mu$m Observations]
{Final Analysis of ELAIS 15 $\mu$m Observations:\\Method, Reduction and Catalogue}
\author[M. Vaccari, C. Lari et al.]
       {M. Vaccari$^{1,2,3}$\thanks{e-mail: m.vaccari@imperial.ac.uk},
        C. Lari$^{4}$,
        L. Angeretti$^{5}$,
        D. Fadda$^{6}$,
        C. Gruppioni$^{7,8}$,
        F. Pozzi$^{5}$,
        \newauthor
        O. Prouton$^{1}$,
        H. Aussel$^{9}$,
        P. Ciliegi$^{8}$,
        A. Franceschini$^{1}$,
        E. Gonz\'alez-Solares$^{10}$,
        \newauthor
        F. La Franca$^{11}$,
        S. Oliver$^{12}$,
        I. Perez-Fournon$^{13}$,
        M. Rowan-Robinson$^{1}$,
        S. Serjeant$^{14}$
        \newauthor
        and
        P. V\"{a}is\"{a}nen$^{15,16}$
        \\\\
        $^{1}$Dipartimento di Astronomia, Universit\`a di Padova,
        Vicolo dell'Osservatorio 2, I-35122, Padova, Italy\\
        $^{2}$CISAS "G. Colombo", Universit\`a di Padova,
        Via Venezia 15, I-35131, Padova, Italy\\
        $^{3}$Astrophysics Group, Blackett Laboratory, Imperial College, 
        Prince Consort Road, London, SW7 2AZ, UK\\
        $^{4}$Istituto di Radioastronomia, CNR,
        Via Gobetti 101, I-40122, Bologna, Italy\\
        $^{5}$Dipartimento di Astronomia, Universit\`a di Bologna,
        Via Ranzani 1, I-40127, Bologna, Italy\\
        $^{6}$Spitzer Science Center, MC 220-6,
        1200 East California Boulevard, Pasadena, CA 91125, USA\\
        $^{7}$Osservatorio Astronomico di Padova, INAF,
        Vicolo dell'Osservatorio 5, I-35122, Padova, Italy\\
        $^{8}$Osservatorio Astronomico di Bologna, INAF,
        Via Ranzani 1, I-40127, Bologna, Italy\\
        $^{9}$Institute for Astronomy, University of Hawaii,
        2680 Woodlawn Drive, Honolulu, HI 96822, USA\\
        $^{10}$Institute of Astronomy, University of Cambridge,
        The Observatories, Madingley Road, Cambridge, CB3 0HA, UK\\
        $^{11}$Dipartimento di Fisica, Universit\`a di "Roma Tre",
        Via della Vasca Navale 84, I-00146, Roma, Italy\\
        $^{12}$Astronomy Centre, Department of Physics \& Astronomy, 
        University of Sussex, Brighton, BN1 9QJ, UK\\
        $^{13}$Instituto de Astrof\'{\i}sica de Canarias,
        Via Lactea S/N, E-38200, La Laguna, Spain\\
        $^{14}$Centre for Astrophysics and Planetary Science,
        School of Physical Sciences, University of Kent,
        Canterbury, Kent, CT2 7NZ, UK\\
        $^{15}$European Southern Observatory, Alonso de Cordova 3107,
        Casilla 19001 Santiago 19, Vitacura, Santiago, Chile\\
        $^{16}$Observatory, P.O. Box 14, FIN-00014,
        University of Helsinki, Finland\\
       }
\date{Accepted 2004 ? ?.
      Received 2004 ? ?;
      in original form 2004 ? ?}
\pagerange{\pageref{firstpage}--\pageref{lastpage}}
\pubyear{2004}
\begin{document}
\maketitle
\label{firstpage}
\begin{abstract}
We present the Final Analysis of the European Large Area ISO Survey (ELAIS)
15 $\mu$m observations, carried out with the ISOCAM instrument on board the
Infrared Space Observatory (ISO).

The data reduction method, known as LARI method, is based on a mathematical
model of the detector's behaviour and was specifically designed for the
detection of faint sources in ISO-CAM/PHOT data. The method is fully
interactive and leads to very reliable and complete source lists.

The resulting catalogue includes 1923 sources detected with $S/N > 5$ in the
\mbox{0.5 -- 100 mJy} flux range and over an area of 10.85 \mbox{deg$^2$}
split into four fields, making it the largest non-serendipitous extragalactic
source catalogue obtained to date from ISO data.

This paper presents the concepts underlying the data reduction method
together with its latest enhancements. The data reduction process, the
production and basic properties of the resulting catalogue are then
discussed.
\end{abstract}
\section{Introduction}
The Infrared Astronomical Satellite
\citep[IRAS,][]{Neugebauer_et_al_1984,Soifer_et_al_1987} was extremely
successful in characterizing for the first time the global properties of the
mid- and far-infrared sky, carrying out an all-sky survey at wavelengths
of 12, 25, 60 and 100 $\mu$m and leading to discoveries such as
those of Luminous, Ultraluminous and Hyperluminous Infrared Galaxies
(LIRGs, ULIRG and HLIRGs, respectively),
a substantial population of evolving starbursts and the detection of
large-scale structure in the galaxy distribution \citep{Saunders_et_al_1991}.

Unfortunately, the IRAS view was typically limited to the very local Universe
($z \stackrel{<}{_\sim} 0.2$), thus hampering statistical studies of infrared-luminous
galaxies at cosmological redshifts.
Only few sources were detected by IRAS at higher redshifts, typically ULIRGs 
magnified by gravitational lenses, like F10214+4724
\citep[$z=2.28$,][]{Rowan-Robinson_et_al_1991}.
In particular only about 1\,000 galaxies were
detected all over the sky in IRAS 12 \mbox{$\mu$m} band.
Infrared source counts based on IRAS data
\citep{Rowan-Robinson_et_al_1984,Soifer_et_al_1984}
showed some marginally significant excess of faint sources with respect
to no evolution models
\citep{Hacking_et_al_1987,Franceschini_et_al_1988,Lonsdale_et_al_1990,Gregorich_et_al_1995,Bertin_et_al_1997},
but not enough statistics and dynamic range in flux to discriminate between
evolutionary scenarios were available.

Although conceived as an observatory-type mission, the Infrared Space
Observatory \citep[ISO,][]{Kessler_et_al_1996} was in many ways the natural
successor to IRAS, bringing a gain of a factor $\sim 1000$ in sensitivity
and $\sim 10$ in angular resolution in the mid-infrared.
A substantial amount of ISO observing time was therefore devoted to field
surveys aimed at detecting faint infrared galaxies down to cosmological
distances.
Such surveys were conceived as complementary in flux depth and areal coverage,
allowing a systematic investigation of the extragalactic sky down to so far
unattainable flux densities at both mid and far infrared wavelengths, whose
results are summarized by \citet{Genzel_and_Cesarsky_2000}.
In particular, extragalactic 15 \mbox{$\mu$m} source counts determined with
ISOCAM \citep{Elbaz_et_al_1999,Gruppioni_et_al_2002} have revealed a
significant departure from Euclidean slope within the 1 - 5 mJy flux range,
which has been interpreted as evidence for a strongly evolving population
of starburst galaxies.

The European Large Area ISO Survey
\citep[ELAIS,][]{Oliver_et_al_2000,Rowan-Robinson_et_al_2003}
was the most ambitious non-serendipitous survey and the largest Open Time
project carried out with ISO, aimed at bridging the flux gap between IRAS
all-sky survey and ISO deeper surveys. ELAIS observations mapped areas of
about 12 \mbox{deg$^2$} at 15 and 90 $\mu$m and smaller areas at 7 and 175
$\mu$m with the ISOCAM \citep[7 and 15 $\mu$m]{Cesarsky_et_al_1996} and
ISOPHOT \citep[90 and 175 $\mu$m]{Lemke_et_al_1996} cameras.
Most importantly, ELAIS 15 $\mu$m observations are the only ones allowing
to sample the 1 - 5 mJy flux range, where most of the source evolution
appears to take place.

Since the project approval, the ELAIS consortium, grown in time to a total
of 76 collaborators from 30 European institutes, has undertaken an extensive
program of ground-based optical and near-infrared imaging and spectroscopy.
Thanks to such an extensive multi-wavelength coverage, the ELAIS fields
have now become among the best studied sky areas of their size, and natural
targets of on-going or planned large-area surveys with the most powerful
ground- and space-based facilities. Further details on ELAIS multi-wavelength
observations and catalogues are presented in \citet{Rowan-Robinson_et_al_2003}.
After the loss of the WIRE satellite, notwithstanding the observations
at several infrared wavelengths soon to come from Spitzer and later from SOFIA
and Herschel, ISO observations will remain a valuable database for many years
to come. In particular, until the advent of JWST, ELAIS 15 \mbox{$\mu$m}
observations will provide a complementary view on three areas (S1, N1 and N2)
which will be covered at different wavelengths as part of the
Spitzer Wide-Area Extragalactic Survey \citep[SWIRE,][]{Lonsdale_et_al_2003}.
Thus the need of reducing such data with the uttermost care and provide
the community with an agreed-upon legacy from the ELAIS project.

This paper presents the Final Analysis of ELAIS 15 \mbox{$\mu$m} observations,
and is structured as follows.
In Section~\ref{elais15.sec} a brief description of the most relevant
aspects of ELAIS 15 $\mu$m dataset is given.
Section~\ref{datared.sec} describes the data reduction method and its
improvements. In Section~\ref{autosim.sec} the technique employed for
flux determination and its results are presented.
Section~\ref{simulations.sec} details the results of the simulations
that were carried out in order to assess the performance of the data
reduction method and thus the quality of the resulting catalogue.
In Sections~\ref{astroacc.sec} and \ref{photoacc.sec}, respectively,
estimates of the achieved astrometric and photometric accuracy are given.
Section~\ref{optids.sec} summarizes the identification of 15 $\mu$m sources
in optical and near-infrared images, while Section~\ref{photocal.sec}
describes the procedure adopted to establish the catalogue photometric
calibration. Finally, Section~\ref{catalogue.sec} describes gives a basic
description of the catalogue contents.
\section{The ELAIS 15 \mbox{$\mu$m} Dataset}\label{elais15.sec}
The ELAIS 15 $\mu$m main\footnote
{Note that smaller sky regions observed with an higher redundancy as part of
the ELAIS project, such as the S2 field (whose data reduction and analysis
is described by \citet{Pozzi_et_al_2003}), the X1,...,6 fields and the
ultra-deep portion of the N1 field, which was observed ten times, are not
considered in this work. See \citet{Oliver_et_al_2000} for further details
on these smaller fields.}
dataset is made up of 28 rasters (ISO basic imaging observations), each
covering an area of about $40^\prime \times 40^\prime$, divided into 4
fields, one (S1) in the southern hemisphere and three (N1, N2 and N3) in the
northern one. 
Small superpositions at the boundaries and a limited degree of
redundancy on portions of the fields give a total covered area of
10.85 \mbox{deg$^2$}.

\begin{figure*}
\centering
\begin{minipage}{0.95\textwidth}
\begin{minipage}{0.475\textwidth}
\resizebox{\textwidth}{!}{\includegraphics*{\figdir/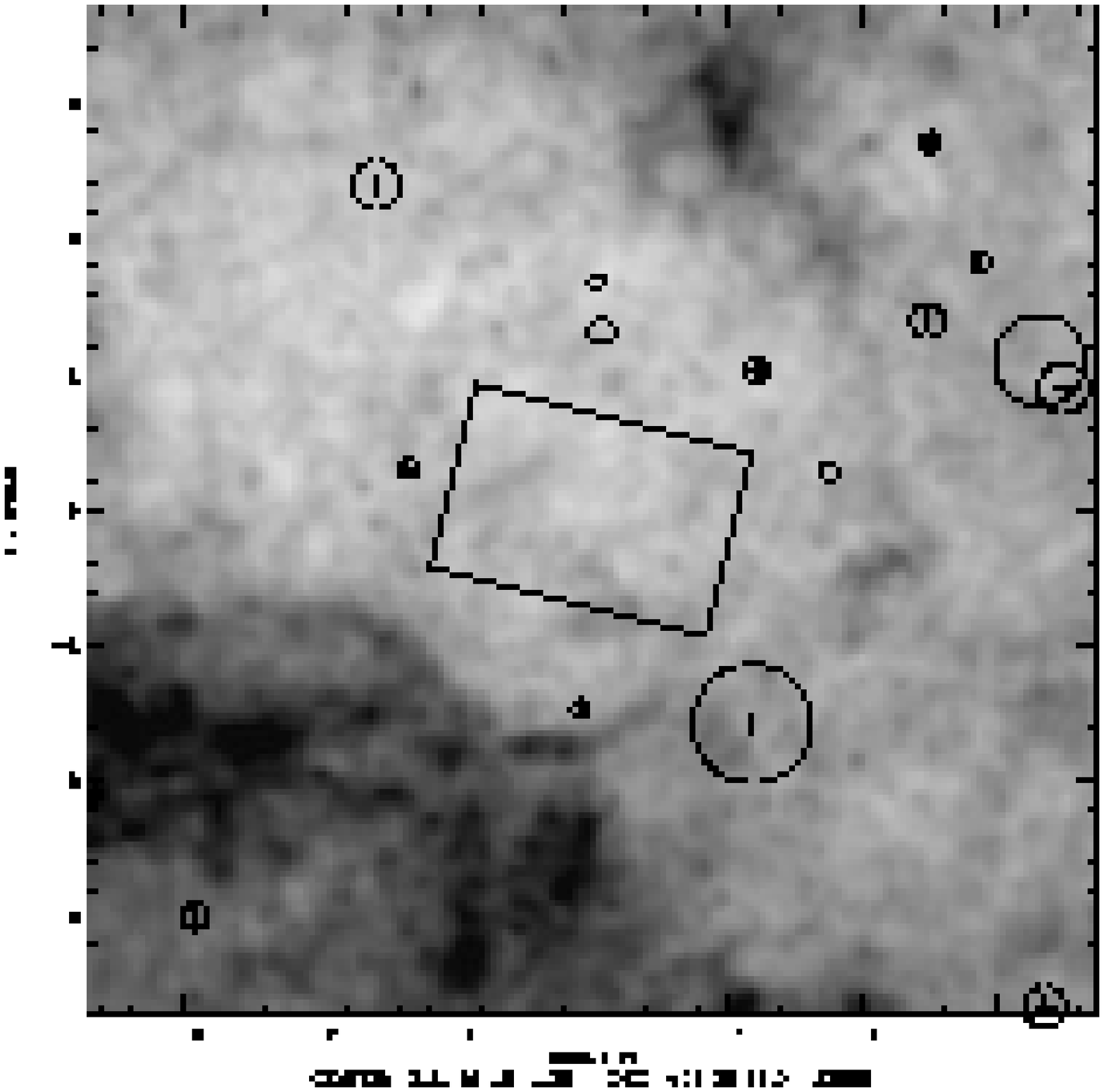}}
\end{minipage}
\hfill
\begin{minipage}{0.475\textwidth}
\resizebox{\textwidth}{!}{\includegraphics*{\figdir/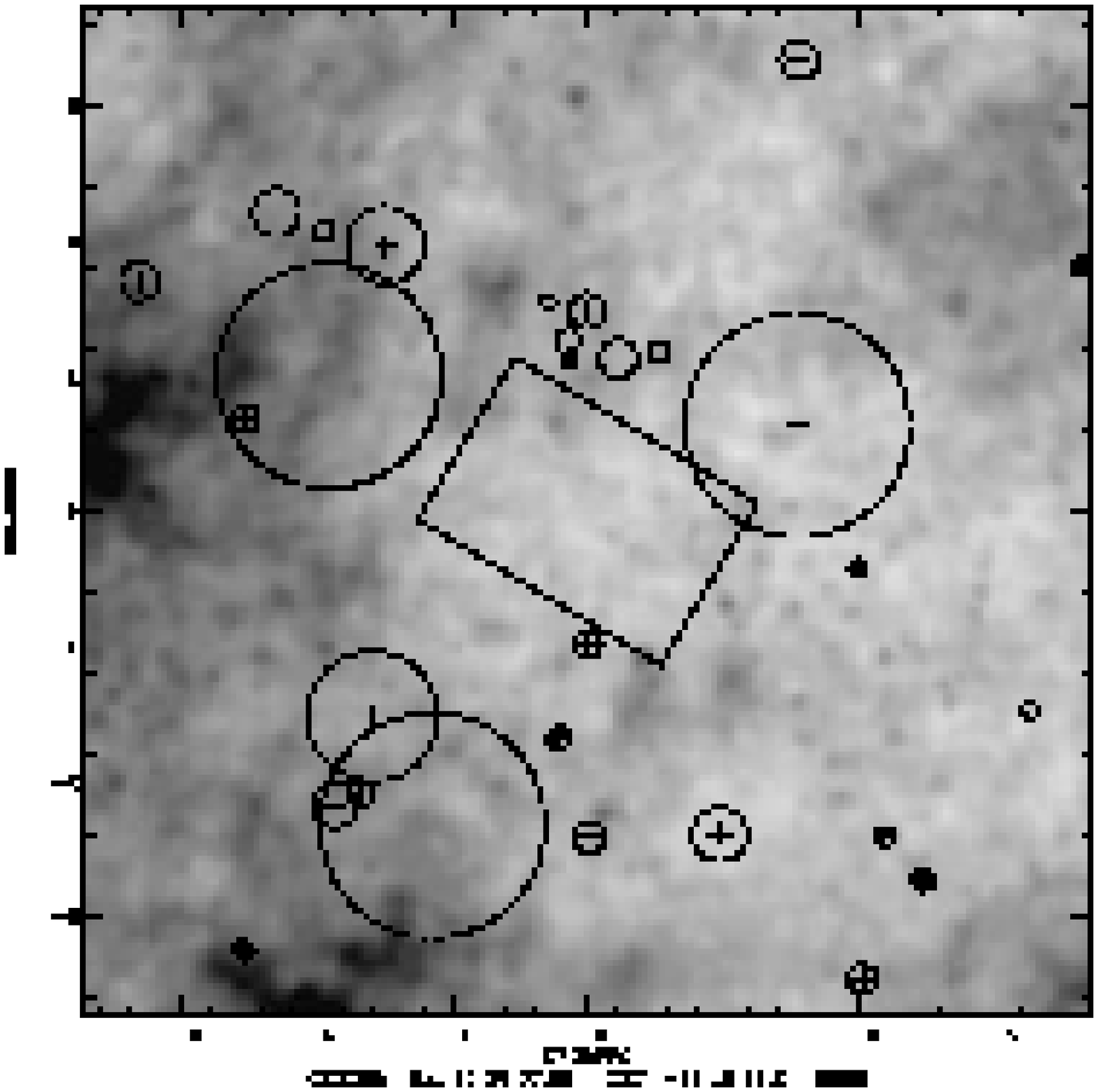}}
\end{minipage}
\end{minipage}
\\\vspace*{0.25cm}
\begin{minipage}{0.95\textwidth}
\begin{minipage}{0.475\textwidth}
\resizebox{\textwidth}{!}{\includegraphics*{\figdir/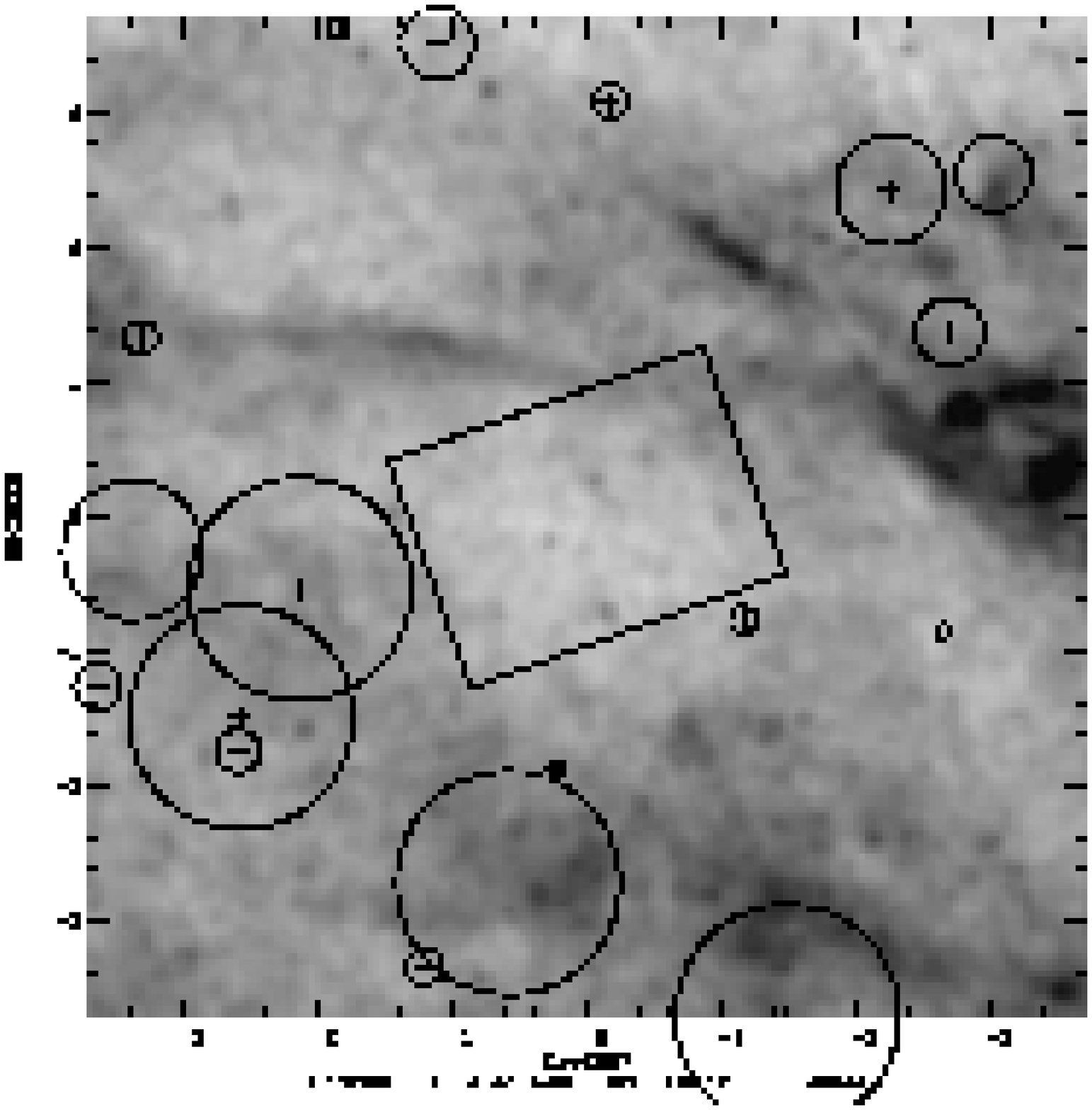}}
\end{minipage}
\hfill
\begin{minipage}{0.475\textwidth}
\resizebox{\textwidth}{!}{\includegraphics*{\figdir/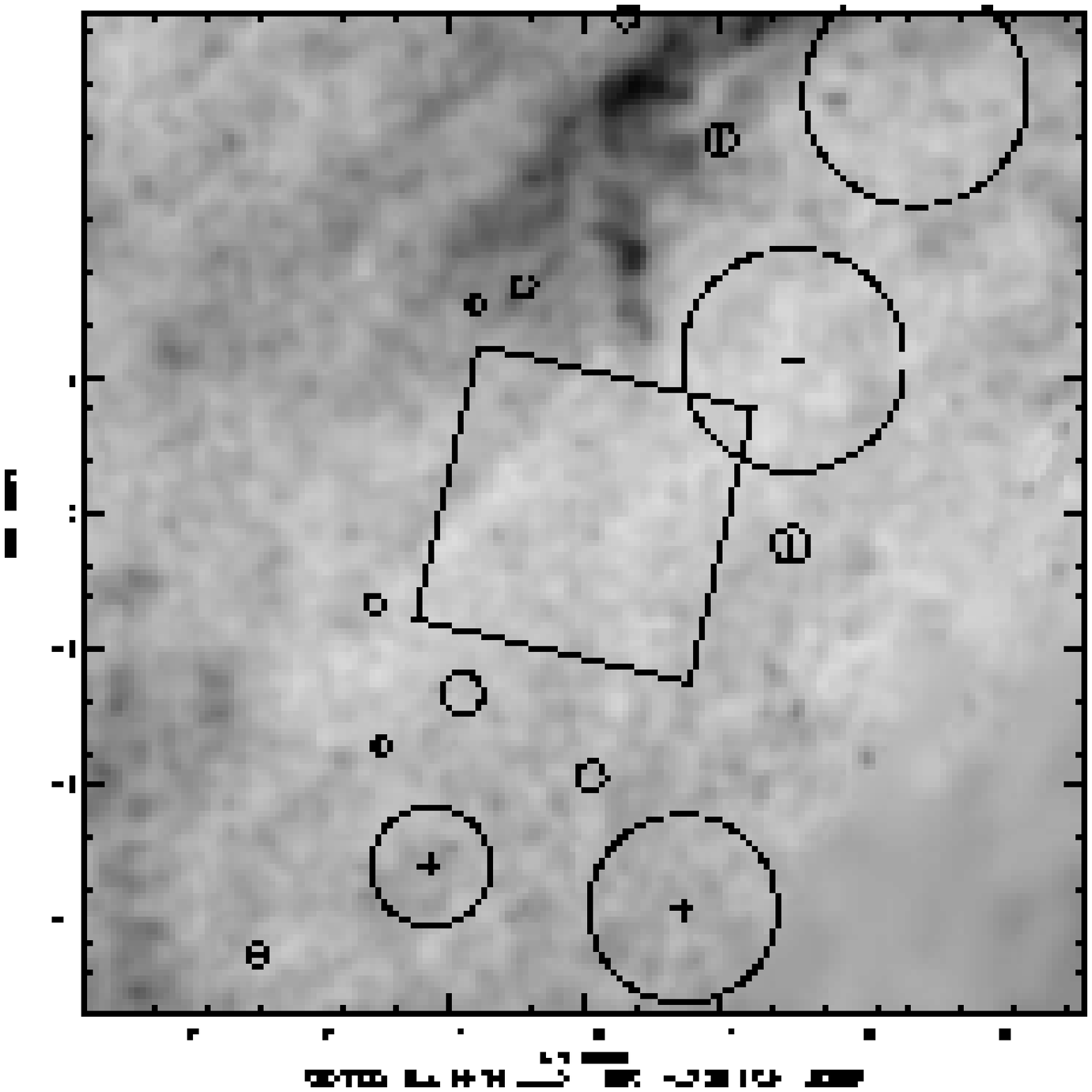}}
\end{minipage}
\end{minipage}
\\
\caption{Sky Locations of ELAIS 15 $\mu$m Fields. From top left to bottom right: N1, N2, N3 and S1 areas. Greyscales indicate COBE normalized IRAS 100 $\mu$m intensity maps from \citet{Schlegel_et_al_1998}. IRAS sources with 12 $\mu$m fluxes brighter than 0.6 Jy are also drawn as circles with radii proportional to their fluxes}
\label{fields.fig}
\end{figure*}

\begin{figure*}
\begin{center}
\centering
\resizebox{0.95\textwidth}{!}{\includegraphics*{\figdir/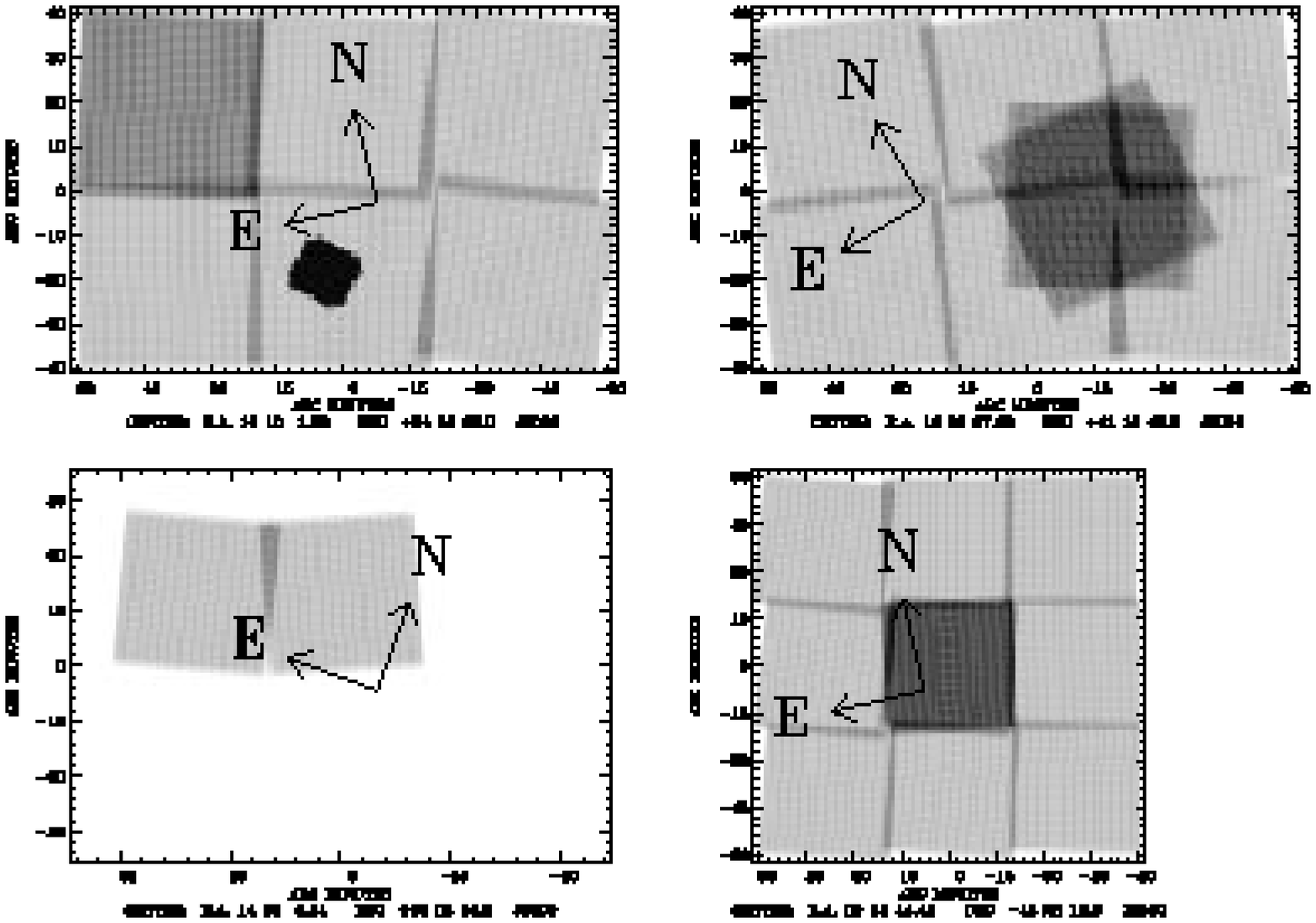}}
\caption{ELAIS 15 $\mu$m sky coverage. From top left to bottom right: N1, N2, N3 and S1 fields. Note that different fields are represented with a slightly different image scale.}
\label{coverage.fig}
\end{center}
\end{figure*}

The fields were selected on the basis of their high ecliptic latitude
($| \beta | > 40^{\circ}$, to reduce the impact of Zodiacal dust emission),
low cirrus emission ($I_{100 \mu m} < 1.5$ \mbox{MJy/sr}) and absence of any
bright ($S_{12 \mu m} > 0.6$ Jy) IRAS 12 $\mu$m source.
In Figure~\ref{fields.fig} the location on the sky of the survey fields
is shown, overlaid on cirrus maps (COBE normalized IRAS maps of
\citet{Schlegel_et_al_1998}). Nearby IRAS sources with 12 $\mu$m fluxes
brighter than \mbox{0.6~Jy} are also plotted.
The overall sky coverage, highlighting the position and redundancy of
single rasters, is illustrated in Figure~\ref{coverage.fig}.

ELAIS 15 $\mu$m observations were carried out operating the ISOCAM instrument
in raster mode using the LW3 filter. In this observing mode, the ISOCAM
$32\times32$ pixel LW detector was stepped across the sky in a grid pattern,
with about half detector width steps in one direction and the whole detector
width steps in the other. Thus, reliability was improved as each sky position
(apart from those  at the boundaries of the raster region) was observed twice
in successive pointings and overheads were reduced because each raster covered
a large area ($\sim40^{\prime} \times 40^{\prime}$).
At each raster pointing (i.e. grid position) the detector was read out
several (typically 10) times, to increase the redundancy in order to be
able to identify cosmic rays impacts and distinguish their severe effects
on the electronics from real sources.
Furthermore, on the raster first pointing, 80 readouts were carried out
to allow the detector to approach stabilization.
Table~\ref{LW3pars.tab} describes the observation parameters for the LW3
observations, while Table~\ref{15mumfields.tab} lists the fields and related
ISO Data Archive filenames making up the dataset.

\begin{table}
\centering
\caption{ELAIS ISOCAM LW3 Observation Parameters}
\label{LW3pars.tab}
\begin{tabular}{lc}
\hline
Parameter & Value \\
\hline
Band $\lambda_{eff}$ & 14.3 $\mu$m \\
Band FWHM Range & 12.0--18.0 $\mu$m \\
Detector gain & 2\addtocounter{footnote}{1}$\fnsymbol{footnote}$\addtocounter{footnote}{-1} \\
Integration time & 2 s \\
Number of exposures per pointing & 10 \\
Number of stabilization exposures & 80 \\
Pixel field of view & 6$^{\prime\prime}$ \\
Number of pixels & 32 $\times$ 32 \\
Number of horizontal and vertical steps & 28 , 14 \\
Number of rasters (including repetitions) & 28 \\
Horizontal and vertical step sizes & 90$^{\prime \prime}$ , 180$^{\prime\prime}$ \\ 
Total area & 10.85 \mbox{deg$^2$} \\
\hline
\end{tabular}
\mbox{\addtocounter{footnote}{1}$\fnsymbol{footnote}$\addtocounter{footnote}{-1}  Except in "test" raster N2\_R\_A where gain was 1}
\end{table}

\begin{table}
\centering
\caption{ELAIS 15 $\mu$m Fields. Field Name, Raster Name, ISO Data Archive (IDA) Official Filename and Raster Centre Coordinates}
\label{15mumfields.tab}
\begin{tabular}{ccccc}
\hline
Field & Raster &   IDA   &  RA (J2000) & Dec (J2000) \\
\hline
S1 & S1\_1    & 23200251 & 00 30 25.40 & -42 57 00.3 \\
   & S1\_2    & 23200353 & 00 31 08.20 & -43 36 14.1 \\
   & S1\_3    & 41300955 & 00 31 51.90 & -44 15 27.0 \\
   & S1\_4\addtocounter{footnote}{1}$\fnsymbol{footnote}$\addtocounter{footnote}{-1} & 23300257\addtocounter{footnote}{1}$\fnsymbol{footnote}$\addtocounter{footnote}{-1} & 00 33 59.40 & -42 49 03.1 \\
   & S1\_5\_A & 23300459 & 00 34 44.40 & -43 28 12.0 \\
   & S1\_5\_B & 77500207 & 00 34 44.40 & -43 28 12.0 \\
   & S1\_5\_C & 78502406 & 00 34 44.40 & -43 28 12.0 \\
   & S1\_6\addtocounter{footnote}{2}$\fnsymbol{footnote}$\addtocounter{footnote}{-2} & 41001161\addtocounter{footnote}{2}$\fnsymbol{footnote}$\addtocounter{footnote}{-2} & 00 35 30.40 & -44 07 19.8 \\
   & S1\_7    & 40800663 & 00 37 32.50 & -42 40 41.2 \\
   & S1\_8    & 40800765 & 00 38 19.60 & -43 19 44.5 \\
   & S1\_9    & 41001867 & 00 39 07.80 & -43 58 46.6 \\
N1 & N1\_1    & 30200101 & 16 15 01.00 & +54 20 41.0 \\
   & N1\_2\_A\addtocounter{footnote}{3}$\fnsymbol{footnote}$\addtocounter{footnote}{-3} & 30400103\addtocounter{footnote}{3}$\fnsymbol{footnote}$\addtocounter{footnote}{-3} & 16 13 57.10 & +54 59 35.9 \\
   & N1\_2\_B & 67200103 & 16 13 57.10 & +54 59 35.9 \\
   & N1\_3    & 30500105 & 16 10 34.90 & +54 11 12.7 \\
   & N1\_4    & 30600107 & 16 09 27.00 & +54 49 58.7 \\
   & N1\_5    & 31000109 & 16 06 10.80 & +54 01 08.0 \\
   & N1\_6    & 30900111 & 16 04 59.00 & +54 39 44.3 \\
N2 & N2\_1    & 50200119 & 16 32 59.80 & +41 13 33.2 \\
   & N2\_2    & 51100131 & 16 34 44.50 & +40 38 45.0 \\
   & N2\_3    & 50000723 & 16 36 05.50 & +41 33 11.8 \\
   & N2\_4    & 50200225 & 16 37 48.90 & +40 58 13.1 \\
   & N2\_5    & 50100727 & 16 39 13.80 & +41 52 31.6 \\
   & N2\_6    & 50200429 & 16 40 55.50 & +41 17 22.7 \\
   & N2\_R\_A & 11600721 & 16 35 45.00 & +41 06 00.0 \\
   & N2\_R\_B & 77900101 & 16 35 45.00 & +41 06 00.0 \\
N3 & N3\_3    & 42500237 & 14 29 38.30 & +33 24 49.6 \\
   & N3\_5    & 43800341 & 14 32 38.20 & +33 11 10.3 \\
\hline
\end{tabular}
\mbox{\addtocounter{footnote}{1}$\fnsymbol{footnote}$\addtocounter{footnote}{-1} Originally flagged as "Telemetry Drops" in observation logs}
\mbox{\addtocounter{footnote}{2}$\fnsymbol{footnote}$\addtocounter{footnote}{-2} Originally flagged as "Unknown Quality" in observation logs}
\mbox{\addtocounter{footnote}{3}$\fnsymbol{footnote}$\addtocounter{footnote}{-3} Originally flagged as "Aborted" in observation logs}
\end{table}
\section{Data Reduction}\label{datared.sec}
Reduction of data obtained with ISO instrumentation has always proved very
difficult for a number of reasons. As far as ISOCAM observations carried out
using its Long Wavelength (LW) detector are concerned, the two most important
instrumental phenomena one has to deal with are the qualitatively very
different effects produced on the detector's electronics by the frequent and
severe cosmic ray impacts, which have long been known and referred to as
\textbf{glitches}, and its sizable transient behaviour after changes in the
incident photon flux, which we will hereafter simply refer to as
\textbf{transients}. In both cases, the cryogenic operational temperatures
of the detector caused it to very slowly respond after these events. Lack of
an accurate modeling of these effects can thus lead to spurious detections
or errors in flux determination.
In ELAIS 15 $\mu$m data reduction, the impact of these effects is increased
by the instrumental parameters which were chosen in order to maximize the
survey area.
The short integration time
($2.1~\mathrm{s} \times 10~\mathrm{frames/pointing}$),
the large raster step (half the detector size along one axis and the whole
detector size along the other one) and the large pixel size (6 arcsec) all
contribute to reduce the redundancy and to increase the undersampling.
Low redundancy and high undersampling, in turn, increase difficulties in
distinguishing sources from strong glitches (low reliability) and
in correctly determining source fluxes (low photometric accuracy).

Roughly speaking, glitches can be divided into three categories according
to the way they shape the detector's output signal, their decay time and
influence on the pixel responsivity: glitches belonging to these different
classes are respectively dubbed \textbf{common glitches}, \textbf{faders}
and \textbf{dippers}.
Slow decreases of the signal following cosmic ray impacts are called faders,
while prominent reductions of the pixel responsivity very slowly recovering
afterwards are called dippers. These two effects are believed to be associated
with proton or $\alpha$ particle impacts on the detector, and have a fairly
long lasting impact on pixel responsivities.
Conversely, the much more frequent impacts of cosmic ray electrons produce
common glitches characterized by a relatively fast decay time, lasting only
a few readouts.
Therefore, the number of frames affected by a single fader or dipper is
much higher than in the case of a common glitch, the pixel responsivity
taking from tens to hundreds of seconds to recover completely.
However, common glitches are much more frequent than faders and dippers and,
may all the same hamper the quality of data reduction.
Thus, all kinds of glitches, if not correctly removed (or, more properly,
corrected for), can lead to spurious detections, or \textbf{unreliability},
and to the loss of genuine sources, or \textbf{incompleteness}.
On the other hand, transients all follow the same pattern, due to the fact
that they arise from the non negligible time it takes for the output signal
to reach the stabilization value after a change in the incident photon flux
has taken place. The measured signal is thus always lower than the true one. 
Failing to model this time effect in data reduction can lead to a systematic
underestimation of source fluxes.
For these reasons, the data cleaning and modeling is an extremely delicate
process requiring great care in order to produce highly reliable sky maps
and source lists.

While it was variously demonstrated that it is possible, at least to a
certain extent, to describe the detector's behaviour adopting some physical
model, the large number of readouts involved in raster observations and the
peculiar nature and strength of noise patterns also require efficient and
robust algorithms to be developed so as to make the actual data reduction
undertaking feasible in a nearly-automatic way.

A number of data reduction methods has thus been developed and tested,
mostly on deep fields (e.g. the PRETI method by \citet{Starck_et_al_1999}
and the Triple Beam Switch method by \citet{Desert_et_al_1999}).
Unfortunately, such methods did not prove completely reliable on shallower
fields, leading to a high number of false detections. Besides, these methods
suffered from the lack of an efficient way to interactively check the quality
of the data reduction when needed. The Preliminary Analysis of ELAIS
15 $\mu$m data \citep{Serjeant_et_al_2000} was thus carried out 
adopting a more traditional approach involving the corroboration of
automated detections through visual inspection by different observers.

The data reduction described in this paper was carried out using the LARI
method \citep{Lari_et_al_2001,Lari_et_al_2003}, a new technique
developed to overcome these difficulties and provide a robust interactive
technique for the reduction and analysis of ISOCAM and
ISOPHOT data, particularly suited for the detection of faint sources and
thus for the full exploitation of their scientific potential.
The method was variously refined, and significantly better results are now
obtained, with respect to the technique used in \citet{Lari_et_al_2001} for
the reduction of the S1 field, so that a thorough re-reduction of all
ELAIS fields seemed appropriate and was thus carried out.
As before, data reduction is carried out within an IDL environment using
mostly purpose-built routines, exploiting CAM Interactive Analysis
\citep[CIA,]{Ott_et_al_2001} software for basic operations only.
%
\subsection{The Model}\label{model.sec}
%
%
The LARI method describes the sequence of readouts, or time history,
of each pixel of ISOCAM LW3 detector in terms of a mathematical model
for the charge release towards the contacts.
Such a model is based on the assumption of the existence, in each pixel,
of two charge reservoirs, a short-lived one $Q_b$, also known as
\textbf{breve}, and a long-lived one $Q_l$, also known as \textbf{lunga},
evolving independently with a different time constant and fed by both
the photon flux and the cosmic rays.
The model is fully charge-conservative, i.e. no decay of accumulated charges
is considered, except towards the contacts, and thus the observed signal $S$
is related to the incident photon flux $I$ and to the accumulated charges
$Q_b$ and $Q_l$ by the
\begin{equation}
S = I - \frac{\ddd Q_{tot}}{\ddd t} = I - \frac{\ddd Q_b}{\ddd t} - \frac{\ddd Q_l}{\ddd t}~,
\end{equation}
where the evolution of these two quantities is governed by the same
differential equation, albeit with a different efficiency $e_i$
and time constant $a_i$
\begin{equation}
\frac{\ddd Q_i}{\ddd t} = e_i\,I - a_i\,Q_{i}^2~~~~~\mathrm{where}~~~i=b,l~,
\end{equation}
so that
\begin{equation}
S = (1-e_b-e_l)\,I + a_b\,Q_b^2 + a_l\,Q_l^2~.
\end{equation}
The values of the parameters $e_i$ and $a_i$ depend on the detector's
physics, and are thus assumed to depend on the exposure time of a given
observation and on the stabilization background level of a given pixel,
according to the relation
\begin{equation}
a_i = \frac{t}{t_0}\,\sqrt{\frac{S}{S_0}}~a_{i,0}~,
\end{equation}
where $a_{i,0}$ is the value of $a_i$ relative to a reference exposure time
$t_0$ and average signal level $S_0$.
In practice, an additive offset signal attributable to thermal dark current
(a component which is otherwise removed in standard dark current subtraction),
is actually added to both $S$ and $I$ in the equation above when it is
estimated to be important, i.e. in the rare cases when the deepest dippers'
depth otherwise exceeds 10\,\% of the stabilization background level.
%
\subsection{Pipeline}
Data reduction begins with a preliminary \textbf{pipeline} incorporating all
necessary steps in order to prepare the data for the temporal fitting
procedure which is the critical step of the LARI method.

Raw data downloaded from the ISO Data Archive are first imported into
the IDL \texttt{raster} structure containing all observational information
using CIA routines.
Likewise, dark current subtraction and conversion from ADU to ADU/gain/s
are carried out using CIA. Then a dedicated IDL structure called
\texttt{liscio} is built in order to contain not only all \texttt{raster}
information but also all ancillary arrays needed in order to carry out
the following reduction.

At this stage potential glitches are identified through a two-step median
filtering process of each pixel time history, or \textbf{deglitching}.
Then a separate routine determined the stabilization background
(or global background, as opposed to the local background described below)
and the aforementioned offset signal.
Such routine was carefully devised to filter out in so far as possible
long-term effects such as stabilization, faders and dippers and thus
provide a reliable estimate of these two parameters the fitting procedure
is particularly sensitive to.
The same routine also identifies the few potential bright sources which
at times the fit failed to recognize by itself, which are then interactively
checked by eye to assess their reliability.
\subsection{Fitting}
The signal as a function of time is finally processed independently for 
each pixel. The fitting procedure models the transients attributable to
changes in incident flux and the features on both short and long time scales
produced by cosmic ray impacts on the time history, modeling glitches as
discontinuities in the charge release.
As seen in Section~\ref{model.sec}, the same values for the $e_i$ and $a_i$
parameters are used for all pixels, apart from the scaling of the $a_i$
according to the exposure time and the stabilization background level,
leaving as free parameters only the charges at the beginning of the
observation and at the "peaks" of glitches. 

Glitches from nearby pixels are also considered when their height is
substantially (i.e. 20 times) higher than the chosen threshold.
The fitting algorithm starts with the strongest potential glitches identified
in pipeline deglitching, assumes discontinuities at these positions and tries
to find a fit to the time history that satisfies the model assumed to describe
the solid-state physics of the detector.
If no acceptable fit is found, the next fainter glitch is considered 
as a potential discontinuity, and so on.
Iteration of the fitting procedure is interrupted when either a satisfactory
(typically 0.2 ADU/gain/s) data-model rms deviation is achieved or the maximum
number of allowed iterations is reached.

At this stage the code estimates several quantities needed to build 
the sky maps on which source extraction will then be performed.
All of these quantities are "recovered" from glitches, i.e. already take
into account discontinuities in charge release assumed to describe glitches
during fit, and their list includes:
\begin{itemize}
\item
the charges stored into the \textbf{breve} and \textbf{lunga} reservoirs
at each readout.
\item
the local background, i.e. the signal to be expected on the basis of the 
previously accumulated charges if only the stabilization background were 
hitting the detector.
\item
the model signal, produced by the incident flux coming from both the 
stabilization background and detected sources.
\item
the \textbf{"unreconstructed"} signal,
i.e. the detected flux recovered from glitches but not from transients.
This is computed as the difference between the measured signal and the
local background.
\item
the \textbf{"reconstructed"} signal,
i.e. the detected flux recovered not only from glitches but also
from transients associated with changes in incident flux.
\end{itemize}
For the sake of clarity, one must emphasize here the differences between
the two kinds of signals (and the corresponding fluxes and sky maps they
will finally turn into) defined above, namely \textbf{unreconstructed}
and \textbf{reconstructed} signals.
Both quantities take into account the effects of the stabilization
background and glitches on the detector, and their values are thus expected
to be negligible if a source is not illuminating the pixel at the given
pointing.
The difference between the two quantities only appears when an additional
signal the code is not able to model otherwise (i.e. as the effect of
a glitch) is detected and attributed
to a source of a given flux. The code then models the transients expected
from such an additional flux and "reconstructs" the signal one would detect
if they did not affect the detector, i.e. if its response were instantaneous.
In other words, unreconstructed signals do not take into account the
effects of this modeling of changes in incident flux, thus representing
the effective charge collected during the exposure, whereas reconstructed
signals recover the charge "loss" due to the slow detector response.
Therefore, the former are systematically lower than the latter.
Actually, simulations show that the code is not actually able to properly
model transients below a certain intensity threshold, thus suggesting
the use of unreconstructed signals only to carry out further processing.

\begin{figure}
\begin{center}
\resizebox{0.9\columnwidth}{0.49\columnwidth}{\includegraphics*{\figdir/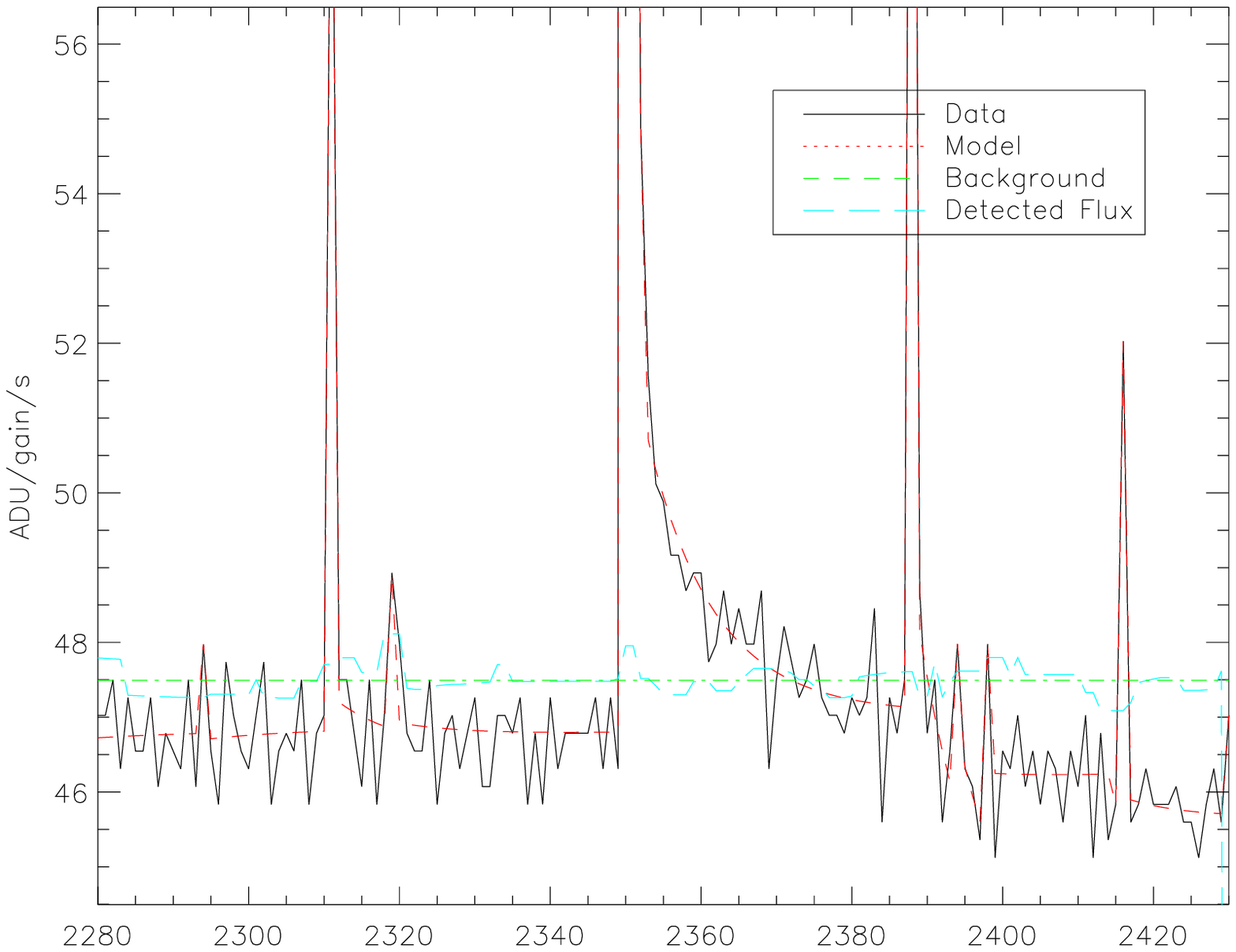}}
\\\textbf{a) Fader}\vspace{0.25cm}\\
\resizebox{0.9\columnwidth}{0.49\columnwidth}{\includegraphics*{\figdir/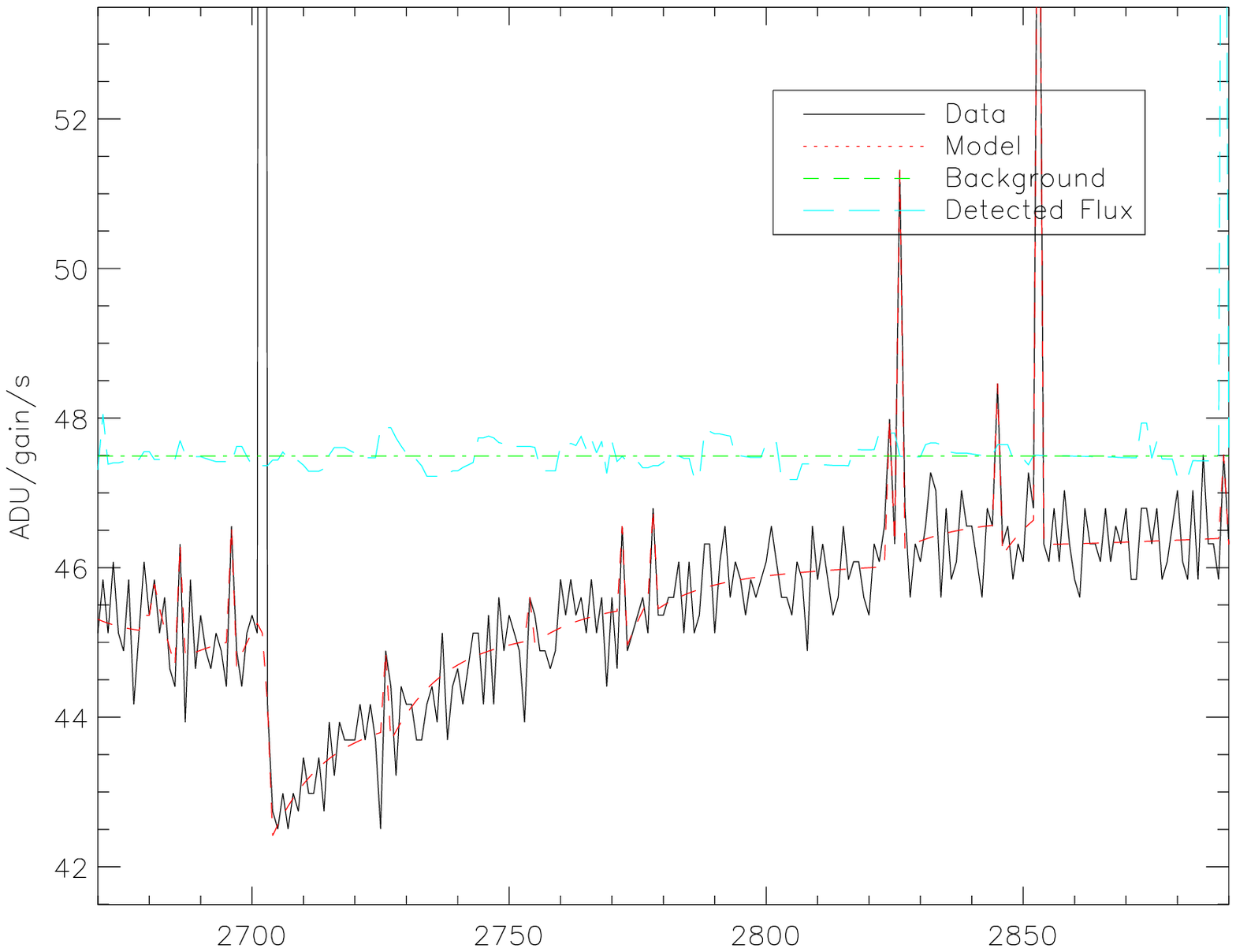}}
\\\textbf{b) Dipper}\vspace{0.25cm}\\
\resizebox{0.9\columnwidth}{0.49\columnwidth}{\includegraphics*{\figdir/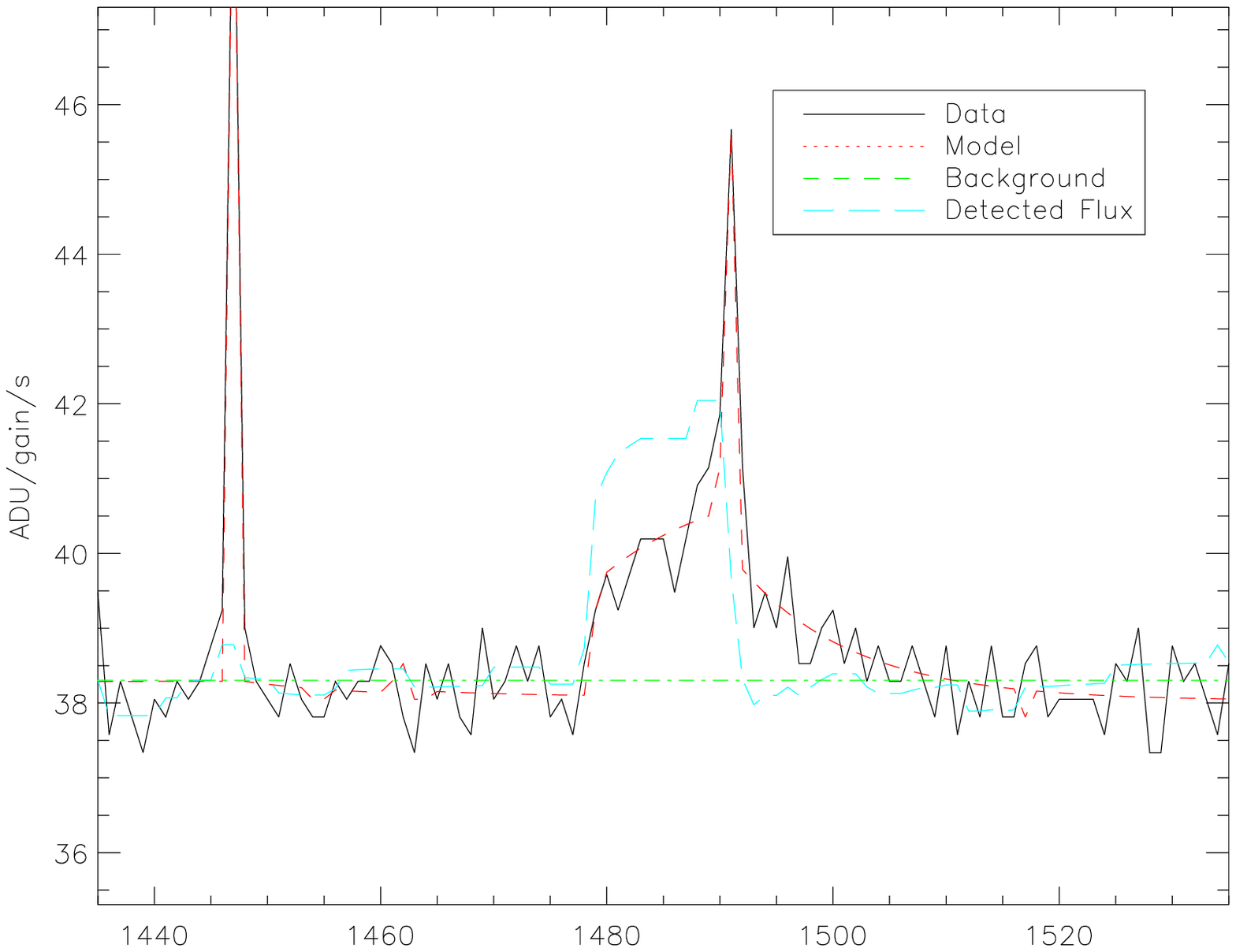}}
\\\textbf{c) Bright source}\vspace{0.25cm}\\
\resizebox{0.9\columnwidth}{0.49\columnwidth}{\includegraphics*{\figdir/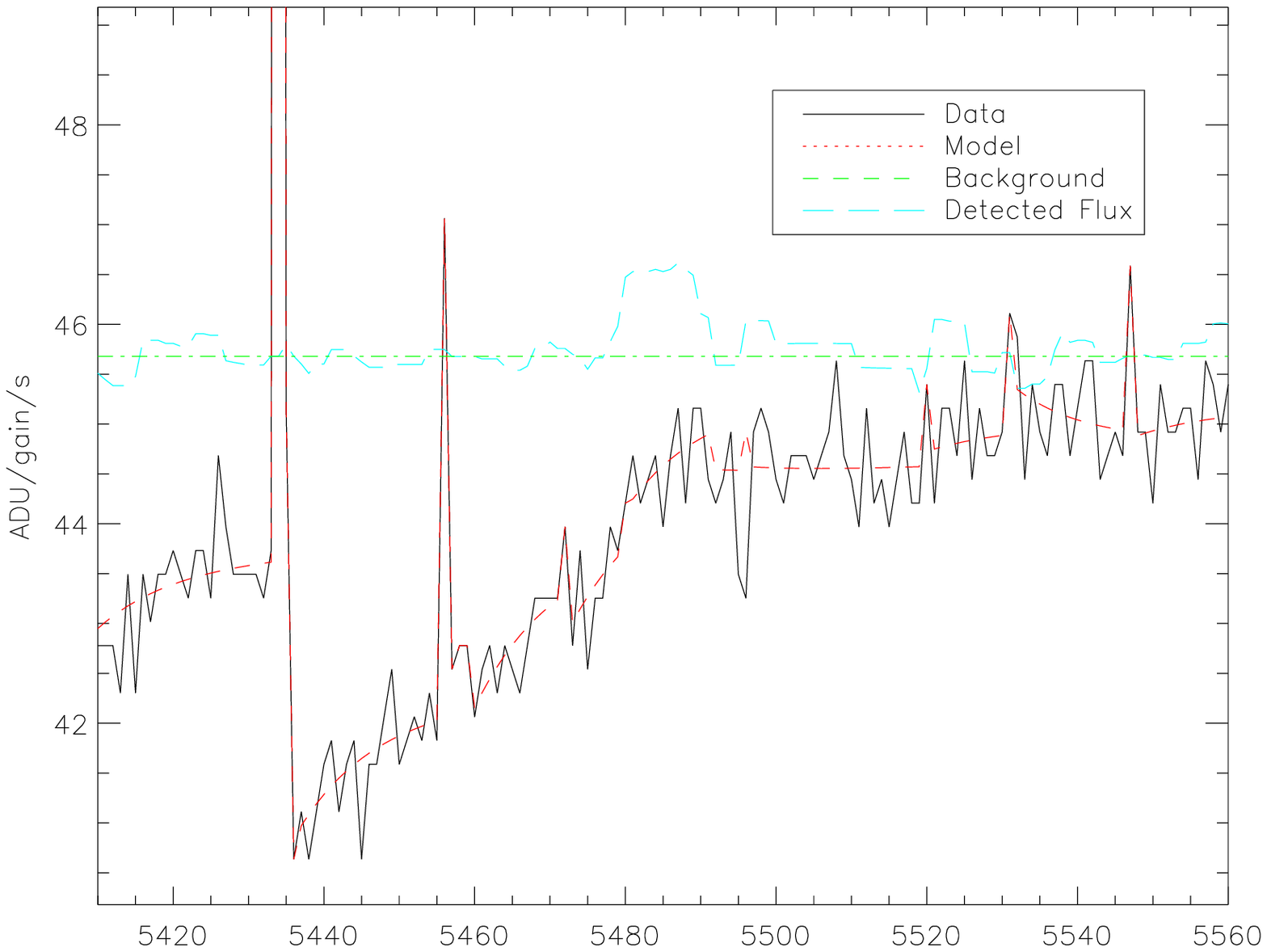}}
\\\textbf{d) Faint source}\vspace{0.25cm}\\
\caption{Different troublesome situations in pixel time histories:
a) Recovery of stabilization background level after a fader b) Recovery of stabilization background level after a dipper c) Detection of a bright source hidden by a strong common glitch d) Faint source hidden by the recovery of the stabilization background level after a dipper. Here "Background" stands for stabilization background and "Detected Flux" for reconstructed flux.}
\label{fits.fig}
\end{center}
\end{figure}

Figure~\ref{fits.fig} shows how a successful fit is able to describe
cosmic-ray-induced violent changes in the signal level and thus recover useful
information (specifically, source fluxes) from the pixel time history.
Panels a) and b) show two examples of how glitches
(a fader and a dipper, respectively) are described as discontinuities
in the signal level slowly recovering towards the stabilization background,
while panels c) and d) show how sources are detected even on the top
of strong glitches.
The solid line represents the observed data, the short-dashed line
the best-fit model and the long-dashed line the detected (unreconstructed)
flux. The dot-dashed line finally represents the stabilization background
%
\subsection{Interactive Analysis}
After running the fitting procedure on the time history of all pixels making
up a raster, the interactive analysis of fitting results is carried out,
looking in detail at portions of the time history that were not well-fit
by the automated analysis.
The details of the interactive analysis process need to be tuned to the
quality of the specific raster under consideration.
In particular, the choice of thresholds in interactive checks is closely
related with observing parameters such as the exposure time but also with
the varying frequency and severeness of cosmic ray impacts.
Generally speaking, more careful checks could be profitably carried out
on intrinsically higher-quality data. On the other hand, it is desirable
to ensure the most uniform data reduction as practicable. It was thus
decided to adopt a common trade-off between the data quality of all rasters
making up our dataset and to apply it to all rasters.

As a first step, fitting failures flagged by substantial data-model
rms deviations
(higher that 0.23 ADU/gain/s)
or negative signals
(lower than -0.6 ADU/gain/s)
are checked.
Then all sizeable signal excesses
(unreconstructed signals higher than 0.5 ADU/gain/s)
are individually inspected.
Whenever the need arises, a further fit extending to the whole pixel time
history or to a smaller portion of it is carried out. Particularly noisy
regions or very strong individual features completely preventing data
reduction are masked.

\begin{figure*}
\begin{center}
\centering
\resizebox{0.95\textwidth}{!}{\includegraphics*{\figdir/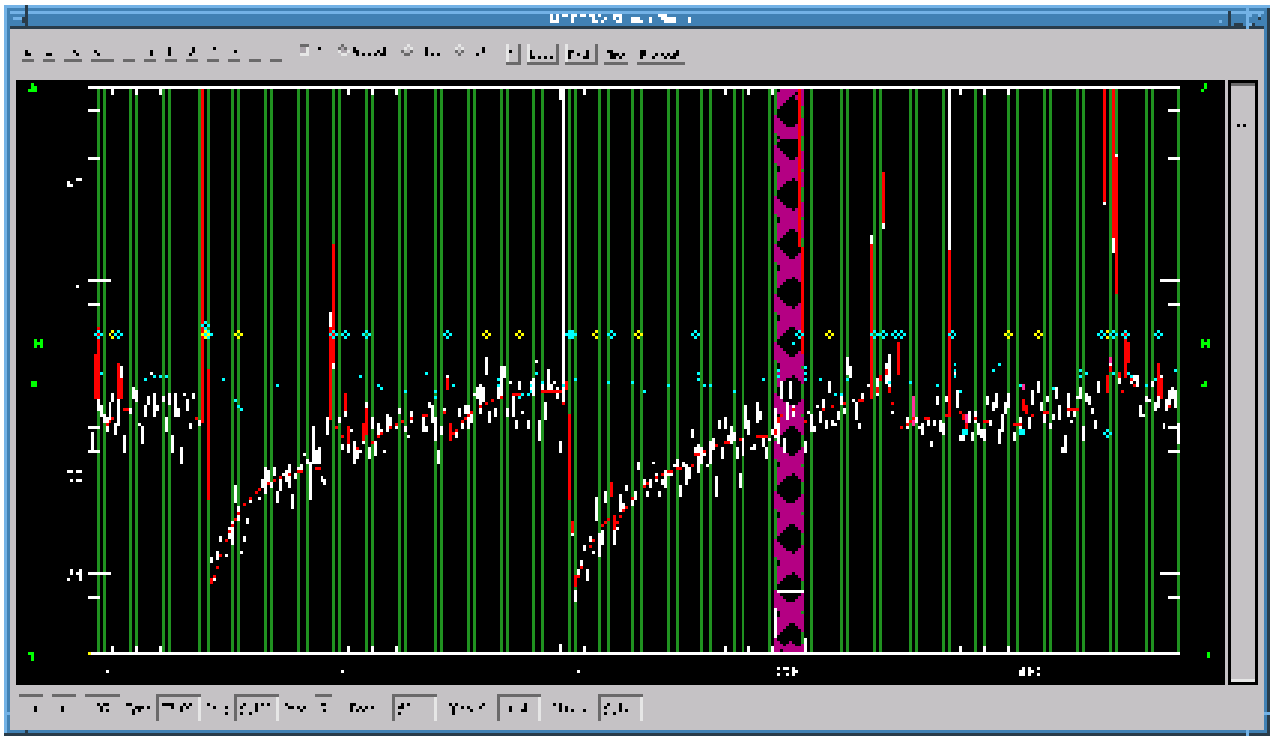}}
\caption{A screenshot of the IDL widget-based Graphical User Interface
used to carry out interactive analysis.}
\label{gui.fig}
\end{center}
\end{figure*}

The massive work of interactive analysis is carried out with an easy-to-use 
IDL widget-based Graphical User Interface, a screenshot of which is shown in
Figure~\ref{gui.fig}, which allows any kind of operation that could be
necessary: data visualization and browsing,  glitch addition and correction,
time history masking and re-fitting. Such software is the main factor
allowing the increase in the volume and quality of the obtained catalogues
with respect to \citet{Lari_et_al_2001}.
\subsection{Mapping}
Once a satisfactory fit is obtained over the whole time history for all pixels,
one can proceed to the generation of sky maps and to source extraction.
After masking glitches and other noisy parts of the time history identified
during Interactive Analysis, flat-fielding is carried out and a signal
estimate for each raster pointing is computed by averaging all readouts
relative to that pointing. The result is converted from ADU/gain/s to mJy
using ISOCAM LW3 standard sensitivity 1 mJy = 1.96 ADU/gain/s.
The "images" thus computed relative to all raster pointings are then projected
onto a $2^{\prime\prime}\times2^{\prime\prime}$ pixel sky map adopting nominal
astrometric information and using a new mapping technique optimized to partly
overcome the severe PSF undersampling due to the large pixel size adopted in
observations.
Here and in the following, all necessary projections
are carried out using the \texttt{projection} C++ code included in CIA.
In so doing, ISOCAM severe field distortion \citep{Okumura_2000} was taken
into account as determined by \citet{Aussel_et_al_1999}, whereas PSF modeling
followed the prescriptions given by \citet{Okumura_1998} for stellar PSFs
but adopting a spectrum of the form $f_{\nu} = constant$, i.e. a closer
match to the expected galaxy spectrum than the Rayleigh-Jeans form used for
stellar spectra. The resulting PSF is larger than the one computed for stars.


Following this procedure, both unreconstructed and reconstructed sky maps
are produced. In both cases, two ancillary maps are also constructed.
The sky coverage (or "NPIX") map contains the number of independent "images"
added together to obtain a pixel value in the final map.
On the other hand, the noise (or "RMS") map contains the estimate of the
root mean square noise of each single pixel in the final map, computed by
scaling the overall root mean square noise of the map as measured at its
centre according to the inverse square root of the noise map.

Source detection is then performed on unreconstructed sky maps.
First, all pixels above a conservatively low flux threshold
(60 $\mu$Jy) are selected, then the IDL Astronomy User's Library
\texttt{find} routine (based on DAOPHOT's homonymous routine, particularly
suited for the detection of point-like sources in crowded fields) is used
to identify positive brightness perturbations around these pixels and return
their peak flux, $S/N$, centroid and shape parameters such as roundness
and sharpness.

At this stage, further interactive checks are performed on all sources
detected with a $S/N$ greater than 5 to assess their reliability.
More specifically, sources detected on the sky map are projected back onto
the time history to identify all raster pointings where the source is
supposed to contribute a significant signal. Then all these pointings
are individually checked and, if necessary, refitted, to improve the fit
and thus recover lost signals or remove spurious ones.
The overall results of these further checks are an increased reliability
and an improved astrometric and photometric accuracy.
\subsection{Mosaicing}\label{mosaicing.sec}
Up to this stage, all rasters are reduced and processed individually.
However, in order to fully exploit the limited redundancy of the
observations, a technique to build a mosaic out of rasters covering
the same field was devised. This is carried out as follows.
Once the reduction of all rasters of interest is completed according to
the above procedure, the necessary corrections to nominal astrometry
are determined as the median offset between ELAIS source positions
expressed with respect to nominal astrometry and the positions of
USNO A2.0 sources found in the field.
This is done through a two-step process. First, the two catalogues are
cross-correlated, assigning to each ELAIS sources its closest USNO
association. The median of the positional differences thus determined,
excluding ELAIS sources with no USNO association within 12 arcsec, is
computed and assumed as a first-order correction to ELAIS source positions.
The cross-correlation procedure is then repeated to calculate a
second-order astrometric correction in exactly the same way, the only
difference being that ELAIS sources with no USNO identification within
3 times the root mean square deviation of ELAIS-USNO association distance
are also excluded during this second step. The sum of the two corrections
is assumed as the raster offset with respect to nominal astrometry, and
the root mean square deviation of ELAIS-USNO distances as the error
in the offset determination. Deviations from nominal astrometry have long
been known to be significant in ISO raster observations, and this is
confirmed by our results, which are summarized in Table~\ref{offsets.tab}.
Most total offsets are greater than or of the order of the pixel field of
view size (6 arcsec), whereas associated errors have a mean value of
0.39 arcsec.
The small errors are not only due to the careful data reduction, but also
to the large number of ELAIS sources, and largely contribute to the very
good overall astrometric accuracy quantified in Section~\ref{astroacc.sec}.

\begin{figure*}
\centering
{\resizebox{0.775\textwidth}{!}{\includegraphics*{\figdir/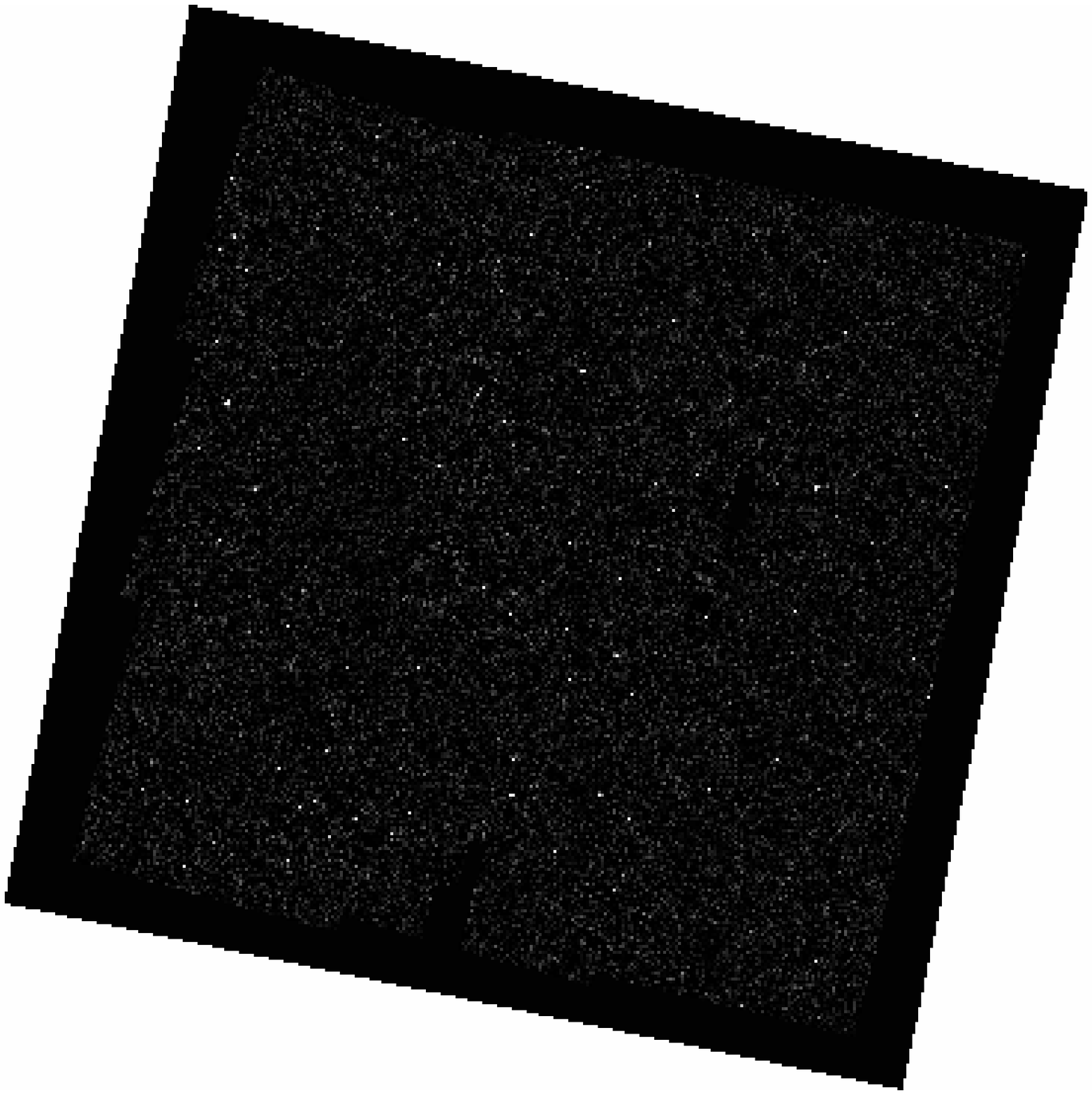}}}
\caption{S1 Field $S/N$ Sky Map. Image size is $2.335 \times 2.335$ deg$^2$.
North is up and East is left.}
\label{s1_map.fig}
\end{figure*}

\begin{figure*}
\centering
{\resizebox{0.5325\textwidth}{!}{\includegraphics*{\figdir/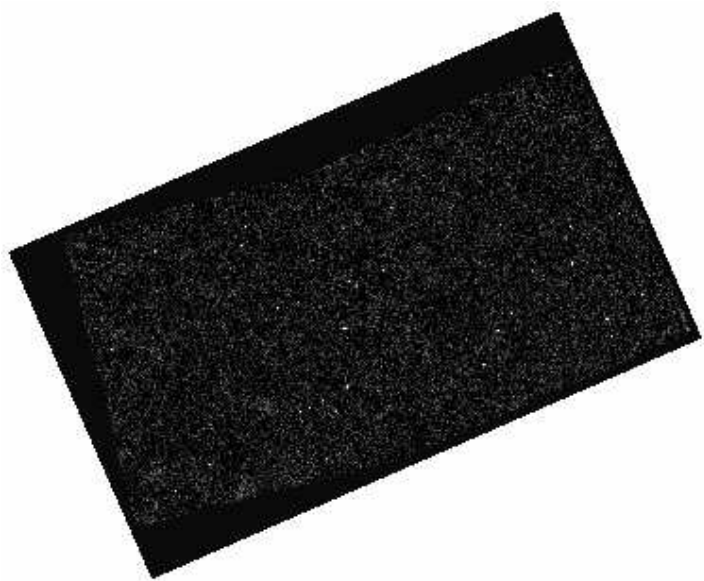}}}
\caption{N3 Field $S/N$ Sky Map. Image size is $0.866 \times 1.454$ deg$^2$.
North is up and East is left.}
\label{n3_map.fig}
\end{figure*}

\begin{figure*}
\centering
{\resizebox{0.675\textwidth}{!}{\includegraphics*{\figdir/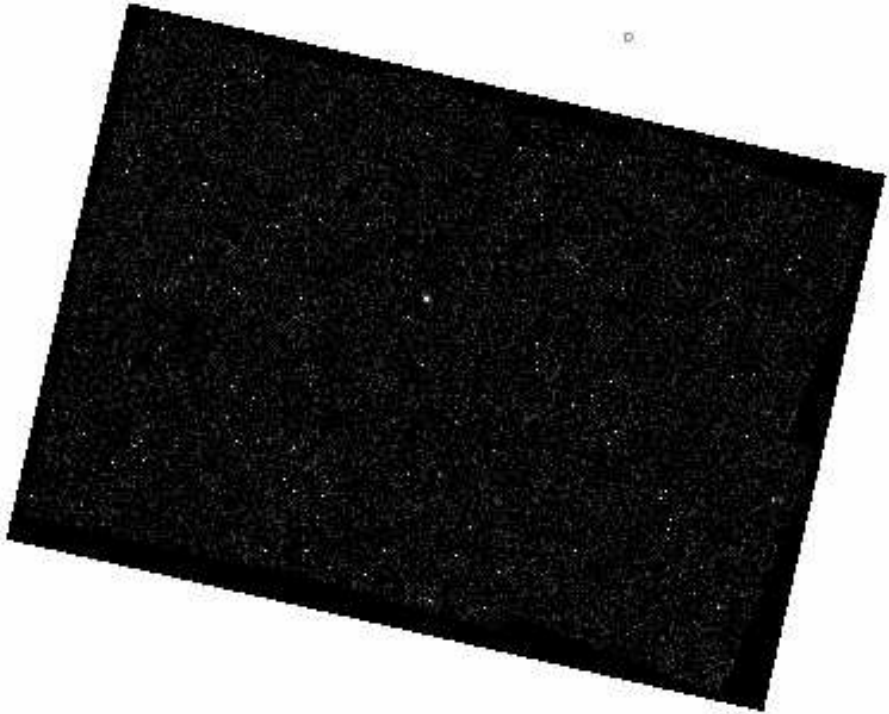}}}
\caption{N1 Field $S/N$ Sky Map. Image size is $1.514 \times 2.134$ deg$^2$.
North is up and East is left.}
\label{n1_map.fig}
\end{figure*}

\begin{figure*}
\centering
{\resizebox{0.7\textwidth}{!}{\includegraphics*{\figdir/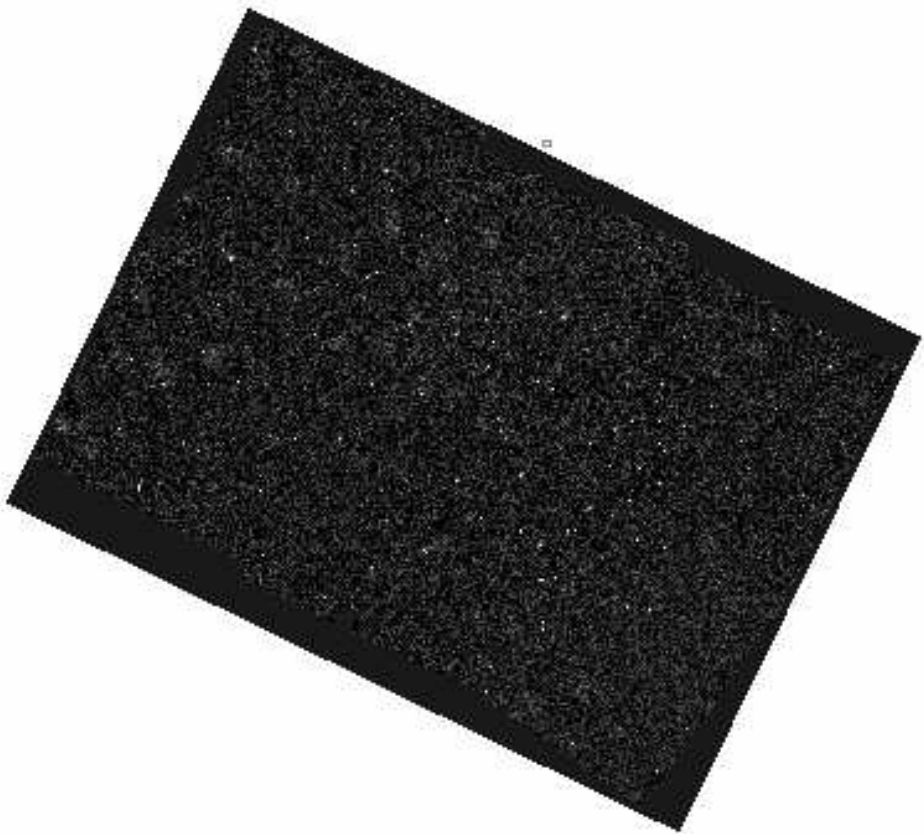}}}
\caption{N2 Field $S/N$ Sky Map. Image size is $1.559 \times 2.117$ deg$^2$.
North is up and East is left.}
\label{n2_map.fig}
\end{figure*}

A common mosaic grid is then built on which different rasters belonging
to the same field are projected taking into account astrometric offsets.
Mapping, source extraction, projection of sources detected with a $S/N$
greater than 5 back on time history and interactive checks are furtherly
performed on this mosaic sky map exactly as on single raster sky maps.
The quality of data reduction is thus improved through cross-checks
of sources detected on different rasters, increasing reliability and
completeness in repeatedly observed regions and partly overcoming the
otherwise severe problems at raster boundaries.
The final $S/N$ sky maps of the four fields are shown in
Figures~\ref{s1_map.fig}, \ref{n3_map.fig}, \ref{n1_map.fig}
and \ref{n2_map.fig}. Typical noise levels are between 20 and 30
\mbox{$\mu$Jy/pixel}.
Such maps, together with noise and sky coverage maps, will soon be
available in FITS format at
\texttt{http://astro.imperial.ac.uk/$\sim$vaccari/elais}.

\begin{table*}
\centering
\caption{Astrometric offsets of ELAIS rasters. Corrections to ISO nominal astrometry as determined through cross-correlation between ELAIS and USNO A2.0 source lists.}
\label{offsets.tab}
\begin{tabular}{ccccccc}
\hline
Raster   & \multicolumn{2}{c}{Nominal Position} & RA ($^{\prime\prime}$) & Dec ($^{\prime\prime}$) & Total ($^{\prime\prime}$) & \\\\
         & RA (J2000) & Dec (J2000) &offset $\pm$ error&offset $\pm$ error&offset $\pm$ error\\
\hline
S1\_1    & 00 30 25.40 & -42 57 00.3 & -2.06 $\pm$ 0.40 & -4.46 $\pm$ 0.38 &  4.91 $\pm$ 0.55 \\
S1\_2    & 00 31 08.20 & -43 36 14.1 & -3.24 $\pm$ 0.22 & +6.86 $\pm$ 0.29 &  7.59 $\pm$ 0.36 \\
S1\_3    & 00 31 51.90 & -44 15 27.0 & +1.57 $\pm$ 0.29 & -7.75 $\pm$ 0.33 &  7.91 $\pm$ 0.44 \\
S1\_4    & 00 33 59.40 & -42 49 03.1 & +0.23 $\pm$ 0.22 & -4.01 $\pm$ 0.27 &  4.02 $\pm$ 0.35 \\
S1\_5\_A & 00 34 44.40 & -43 28 12.0 & -3.50 $\pm$ 0.23 & +9.63 $\pm$ 0.27 & 10.24 $\pm$ 0.35 \\
S1\_5\_B & 00 34 44.40 & -43 28 12.0 & -0.52 $\pm$ 0.21 & -8.10 $\pm$ 0.26 &  8.12 $\pm$ 0.33 \\
S1\_5\_C & 00 34 44.40 & -43 28 12.0 & -3.04 $\pm$ 0.24 & +5.34 $\pm$ 0.29 &  6.14 $\pm$ 0.38 \\
S1\_6    & 00 35 30.40 & -44 07 19.8 & +0.60 $\pm$ 0.43 & -7.14 $\pm$ 0.24 &  7.17 $\pm$ 0.49 \\
S1\_7    & 00 37 32.50 & -42 40 41.2 & +1.26 $\pm$ 0.22 & -5.62 $\pm$ 0.38 &  5.76 $\pm$ 0.44 \\
S1\_8    & 00 38 19.60 & -43 19 44.5 & +0.72 $\pm$ 0.22 & -5.31 $\pm$ 0.24 &  5.36 $\pm$ 0.36 \\
S1\_9    & 00 39 07.80 & -43 58 46.6 & -2.34 $\pm$ 0.19 & +4.61 $\pm$ 0.23 &  5.17 $\pm$ 0.30 \\
N1\_1    & 16 15 01.00 & +54 20 41.0 & +1.96 $\pm$ 0.34 & -9.34 $\pm$ 0.21 &  9.54 $\pm$ 0.40 \\
N1\_2\_A & 16 13 57.10 & +54 59 35.9 & +3.27 $\pm$ 0.31 & -9.48 $\pm$ 0.22 & 10.02 $\pm$ 0.38 \\
N1\_2\_B & 16 13 57.10 & +54 59 35.9 & +4.17 $\pm$ 0.23 & -7.89 $\pm$ 0.15 &  8.92 $\pm$ 0.27 \\
N1\_3    & 16 10 34.90 & +54 11 12.7 & -1.04 $\pm$ 0.42 & -6.18 $\pm$ 0.25 &  6.27 $\pm$ 0.49 \\
N1\_4    & 16 09 27.00 & +54 49 58.7 & +4.60 $\pm$ 0.44 & -6.71 $\pm$ 0.23 &  8.14 $\pm$ 0.50 \\
N1\_5    & 16 06 10.80 & +54 01 08.0 & -8.90 $\pm$ 0.27 & +3.77 $\pm$ 0.18 &  9.67 $\pm$ 0.32 \\
N1\_6    & 16 04 59.00 & +54 39 44.3 & -8.71 $\pm$ 0.32 & +3.88 $\pm$ 0.22 &  9.54 $\pm$ 0.39 \\
N2\_1    & 16 32 59.80 & +41 13 33.2 & -4.68 $\pm$ 0.23 & +7.18 $\pm$ 0.21 &  8.57 $\pm$ 0.31 \\
N2\_2    & 16 34 44.50 & +40 38 45.0 & +2.93 $\pm$ 0.27 & -2.87 $\pm$ 0.27 &  4.10 $\pm$ 0.38 \\
N2\_3    & 16 36 05.50 & +41 33 11.8 & -5.50 $\pm$ 0.46 & +7.30 $\pm$ 0.37 &  9.14 $\pm$ 0.59 \\
N2\_4    & 16 37 48.90 & +40 58 13.1 & -5.36 $\pm$ 0.24 & +7.46 $\pm$ 0.21 &  9.19 $\pm$ 0.32 \\
N2\_5    & 16 39 13.80 & +41 52 31.6 & -5.43 $\pm$ 0.29 & +7.32 $\pm$ 0.24 &  9.11 $\pm$ 0.38 \\
N2\_6    & 16 40 55.50 & +41 17 22.7 & -2.45 $\pm$ 0.28 & +2.32 $\pm$ 0.24 &  3.38 $\pm$ 0.37 \\
N2\_R\_A & 16 35 45.00 & +41 06 00.0 & +1.29 $\pm$ 0.33 & +6.01 $\pm$ 0.21 &  6.15 $\pm$ 0.39 \\
N2\_R\_B & 16 35 45.00 & +41 06 00.0 & +5.91 $\pm$ 0.34 & +4.24 $\pm$ 0.22 &  7.27 $\pm$ 0.40 \\
N3\_3    & 14 29 38.30 & +33 24 49.6 & -1.39 $\pm$ 0.30 & -3.44 $\pm$ 0.16 &  3.71 $\pm$ 0.34 \\
N3\_5    & 14 32 38.20 & +33 11 10.3 & -0.17 $\pm$ 0.31 & -3.52 $\pm$ 0.24 &  3.52 $\pm$ 0.39 \\
\hline
\end{tabular}
\end{table*}
\section{Autosimulation}\label{autosim.sec}
Even after a very careful reduction, flux determination is a very delicate
process owing to the interplay of different factors, the most important
being mapping effects related to PSF undersampling, detector's transients
and other trends due to the limits of the adopted reduction method.
Mapping effects are particularly tricky, since they determine the way one
obtain source total fluxes from corresponding peak fluxes measured on sky
maps by source detection software.
Clearly, the conversion factor between these two quantities is very much
dependent on the actual source position with respect to the centre of the
pixel, a source total flux being nearer to its peak flux the nearer the
source is to the center of the sky map pixel.
The technique we developed, which we called \textbf{autosimulation},
was conceived to model and insofar as possible take into account mapping
and data reduction effects on flux determination by first constructing both
real and theoretical (i.e. noise-free) sky maps, and then use the latter
ones to correct fluxes obtained from the former ones.
Autosimulation consists in simulating sources, on top of noise-free maps,
at each position where a source was actually detected (or, for simulations
described in Section~\ref{simulations.sec}, at their randomly generated
positions), calculating the ratio between peak and total flux for the
simulated source on such theoretical sky maps, and converting the measured
peak flux of the real source to its total flux using such ratio.
Source simulation is carried out in a straightforward way using the same
mathematical model the LARI method is based on, thus also allowing to verify
its predictions.

In order to describe how autosimulation works in some greater detail, a few
definitions and relations which will also be useful to discuss the results
of simulations described in Section~\ref{simulations.sec} must be summarized:
\begin{itemize}
\item $f_s$ is the measured peak flux obtained from real
(i.e. containing glitches, noise and transients) sky maps.
Its value therefore depends on mapping effects, transients and
the adopted data reduction method;
\item $f_0$ is the "theoretical" peak flux obtained from simulated 
(i.e. containing neither glitches nor noise but taking into account
source transients) maps. Its value depends on mapping effects
and transients only.
\item $f_{sr}$ and $f_{0r}$ are analogous to $f_s$ and $f_0$
but are built from reconstructed maps, thus recovering the effects
of transients. As already mentioned, this strictly holds above
a certain flux threshold only, where the values of these quantities
actually cease to depend on transients.
%
%
\end{itemize}
The correction factor for mapping effects is computed as follows.
A simulated source is generated on each position where a real source was
actually detected, assuming a total flux
$S_0 = f_s / \left< f_s / S \right>_{sim}$
based on the measured peak flux $f_s$ and on the average
peak flux / total flux ratio $\left< f_s / S \right>_{sim} = 0.216$
obtained from simulations (see Section~\ref{simulations.sec} and
Figure~\ref{flussivsfs.fig}).
Then the flux estimate corrected for mapping effects is computed as
$S = (f_s / f_0)~S_0$. Source simulation is carried out using the
same model for charge release adopted in fitting the data.
In so doing, however, autosimulation does not correct for more subtle
and elusive effects arising from data reduction, such as flux loss due
to bad fits, which can only be assessed through simulations.
This is described in Section~\ref{fluxdet.sec}.

Such a procedure provides reliable flux estimates which are well compatible
with estimates obtained through conventional aperture photometry for most
sources, but is clearly non-optimal when dealing with extended sources.
For these, as well as for a few very close or blended sources, aperture
photometry usually provide a better estimate of the source flux.
After correcting fluxes for mapping effects, potentially extended sources
are therefore identified through both visual inspection of sky maps
and the calculation of parameters connected to source extension.
Sources flagged by these criteria (totalling 67 out of 1923 making up the
catalogue, or 3.5\,\%) were then individually treated, aperture photometry
with a suitable aperture radius was carried out and the result was adopted
as their most reliable flux estimate.
As to the overall performance of the autosimulation process when compared
with aperture photometry,
autosimulated fluxes and their counterparts calculated through aperture
photometry are compared for all sources in Section~\ref{autoaper.sec}.
\section{Simulations}\label{simulations.sec}
Due to the peculiar nature of ISO data and of the reduction method employed,
it is important to carefully test the performance of the latter on "ideal"
data and sources. In particular, systematic effects on flux estimates related
to the data reduction method can only be probed by these means.
Due to the strong peculiarities of our dataset, which is
characterized by several noise features on different time scales, only
real data can effectively be taken as representative of instrumental behaviour.
Therefore, the effects of artificial sources must be somehow simulated on the
top of real pixel time histories, and data reduction must then be carried out
exactly as done for real sources. Source confusion in the field is thus
slightly increased, but this effect is not critical for ELAIS data.
The LARI Method is based on a physical model of the detecto's behaviour,
so can make straightforward predictions of the effect on the detector of the
additional photons from simulated sources.
On this basis, an extensive set of simulations was carried out to assess
the effects on flux estimates and the overall performance of data reduction
in a statistically meaningful way. The S1 field was chosen for this purpose
as the most representative, being the largest and including both regions
observed only once and repeatedly observed regions (hereafter,
\textbf{non-repeated} and \textbf{repeated} regions, respectively), therefore
allowing to assess the differences in performance warranted by higher redundancy.

Simulations were carried out as follows. First a set of $N=600$ random sky
positions (excluding regions near to real sources, mosaic sky map boundaries
and masked regions of mosaic sky map, i.e. preventing simulated sources
to appear nearer than 15 arcsec to any of these regions) was generated.
Then a logarithmically uniform flux distribution covering the range between
a lower limit $S_l = 0.5~\mathrm{mJy}$ and an upper limit
$S_u = 4.0~\mathrm{mJy}$
\begin{equation}
S_n=S_l \left(\frac{S_u}{S_l}\right)^{n/N}~\mathrm{for}~n=0,N-1
\end{equation}
was coupled with the random positions above to characterize the set of
simulated sources. Following the same procedure, 100 more sources were
simulated (adopting a flux distribution following the same analytical
formula but covering the 0.35--3.5 mJy flux range) in the repeated
regions of the field, thus increasing the otherwise low number of
simulated sources in these regions.
All 700 sources were then projected from their sky position back onto
the pixel time history, simulating their effects as superposed on glitches
and noise characterizing the real data.
In so doing, all portions of time history where simulated sources were
predicted to give a sizeable additional signal were identified.
The fitting procedure was thus re-run on these regions only, and all steps
of data reduction and flux determination were then carried out on simulated
sources exactly as described in Section~\ref{datared.sec} for real sources.
At this stage the positions and fluxes of simulated sources that had been
detected with a $S/N$ greater than 5 were compared with the input values
to calibrate and evaluate the performance provided by our data reduction
method in different respects. The overall number of simulated and detected
sources is given in Table~\ref{simbasic.tab}, whereas the following
Sections detail all relevant aspects of simulations, apart from the
completeness estimates which will be described by \citet{Lari_et_al_2004}
in order to obtain extragalactic source counts.
\begin{table}
\centering
\caption{Number of Simulated and Detected Sources. Total figures are given
together with those for non-repeated and repeated regions.}
\label{simbasic.tab}
\begin{tabular}{lccc}
\hline
& Non-Repeated & Repeated & Total\\
\hline
Simulated Sources & 502 & 198 & 700 \\
Detected Sources  & 230 & 125 & 355 \\
\hline
\end{tabular}
\end{table}
\subsection{Flux Determination}~\label{fluxdet.sec}
The autosimulation process we use for flux determination, though
relatively simple in principle, leads to several systematic effects
which need to be carefully taken into account in order to understand
how it can provide us with the best estimate of source fluxes.
The analysis of its results on simulated sources will also allow us
to test the goodness of our model and the reliability of the related
signal reconstruction process.
As described in Section~\ref{autosim.sec}, autosimulation involves
simulating theoretical (i.e. noise-free) maps containing all
(i.e. both real and simulated) sources of interest, then carrying out
signal reconstruction on both theoretical and real sky maps, thus
determining the theoretical ratio between peak and total fluxes,
and finally applying results determined on theoretical maps to correct
real fluxes.

\begin{figure}
\begin{center}
\centering
\resizebox{0.95\columnwidth}{!}{\includegraphics*{\figdir/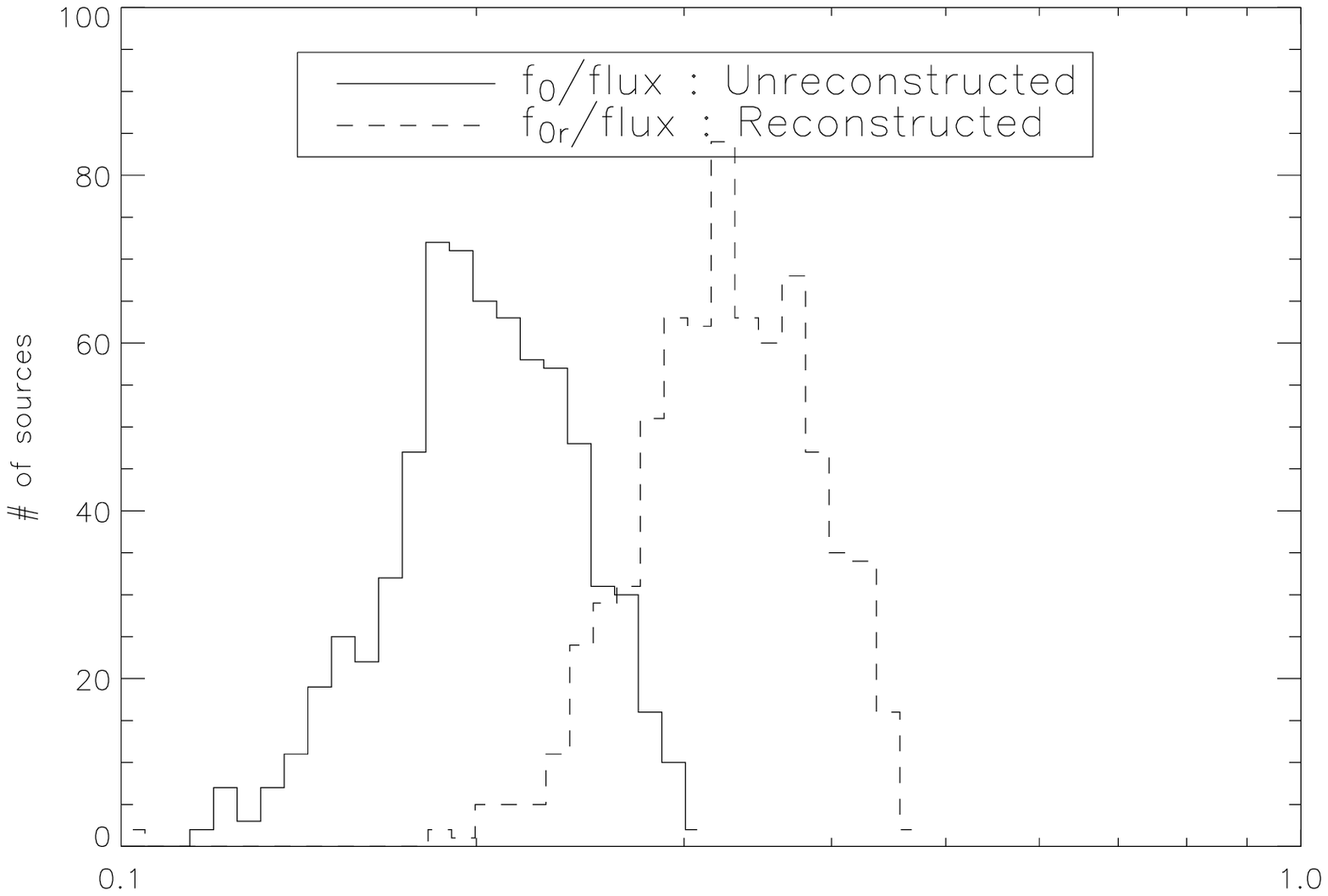}}
\resizebox{0.95\columnwidth}{!}{\includegraphics*{\figdir/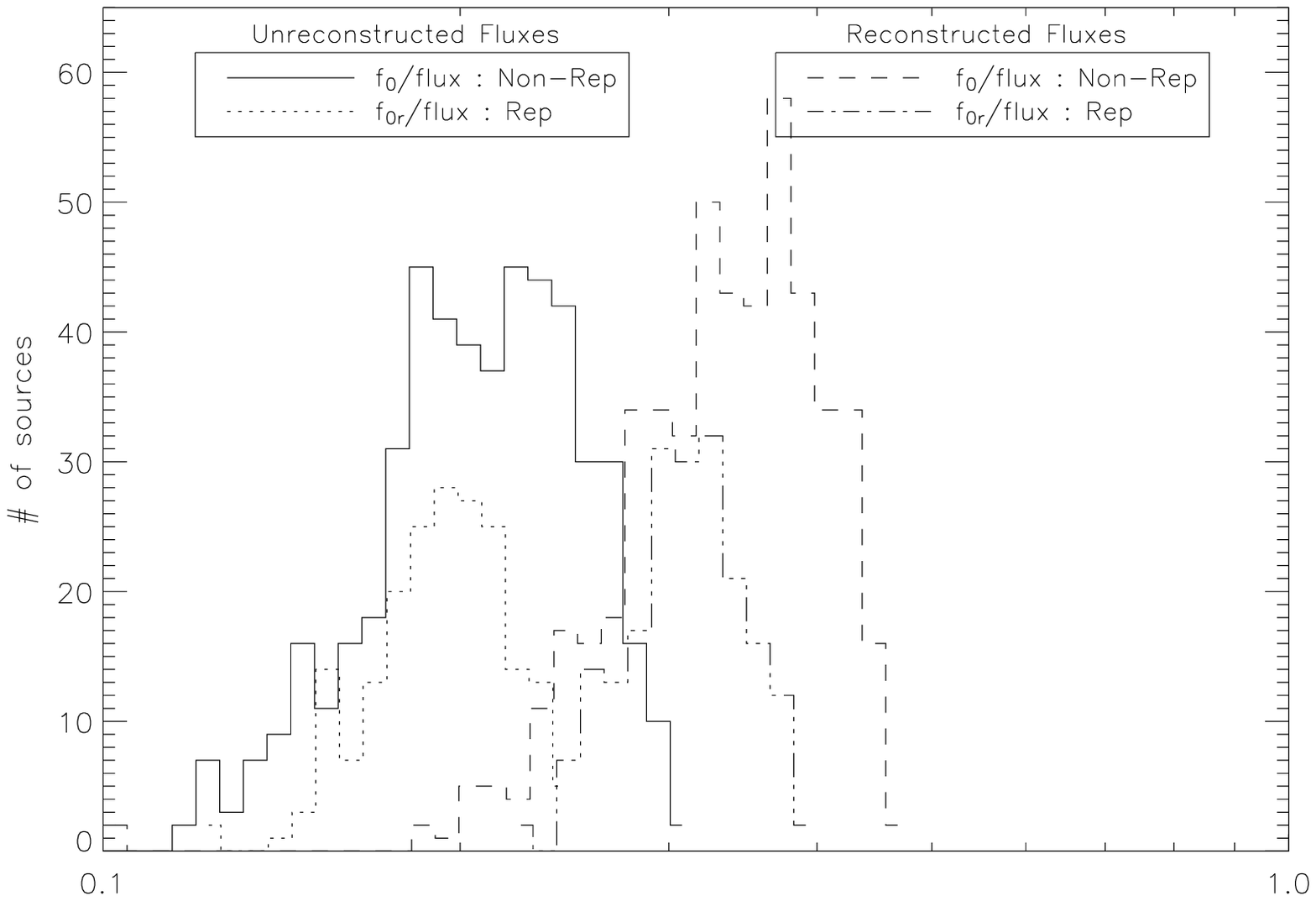}}
\caption{Peak/Total Flux Ratio for theoretical sky maps. Histogram distribution
for both Unreconstructed and Reconstructed maps are given.
Overall distributions are shown (top panel) together with those for
non-repeated and repeated regions separately (bottom panel).}
\label{f0andf0roverflux.fig}
\end{center}
\end{figure}

Figure~\ref{f0andf0roverflux.fig} shows how mapping effects affect
the relation between peak fluxes and total fluxes for both
reconstructed and unreconstructed theoretical sky maps.
Following the nomenclature adopted in Section~\ref{autosim.sec},
the histogram distribution of $f_0 / S_0$ and $f_{0r} / S_0$ ratios
are plotted for all sources (top panel) and separately for non-repeated
and repeated regions (bottom panel). While the two distributions for
non-repeated and repeated regions peak at slightly different values and
thus have a (marginally) smaller width, the $f_{0r}/f_{0}$ ratio is
remarkably similar, 1.61 being its mean value. This means that,
according to our model, due to the short integration time adopted in
observations with respect to the detector's response time constant,
the measured flux, even in a noise-free image, will only equal 60\,\%
of the incident flux.

\begin{figure}
\begin{center}
\centering
\resizebox{0.95\columnwidth}{!}{\includegraphics*{\figdir/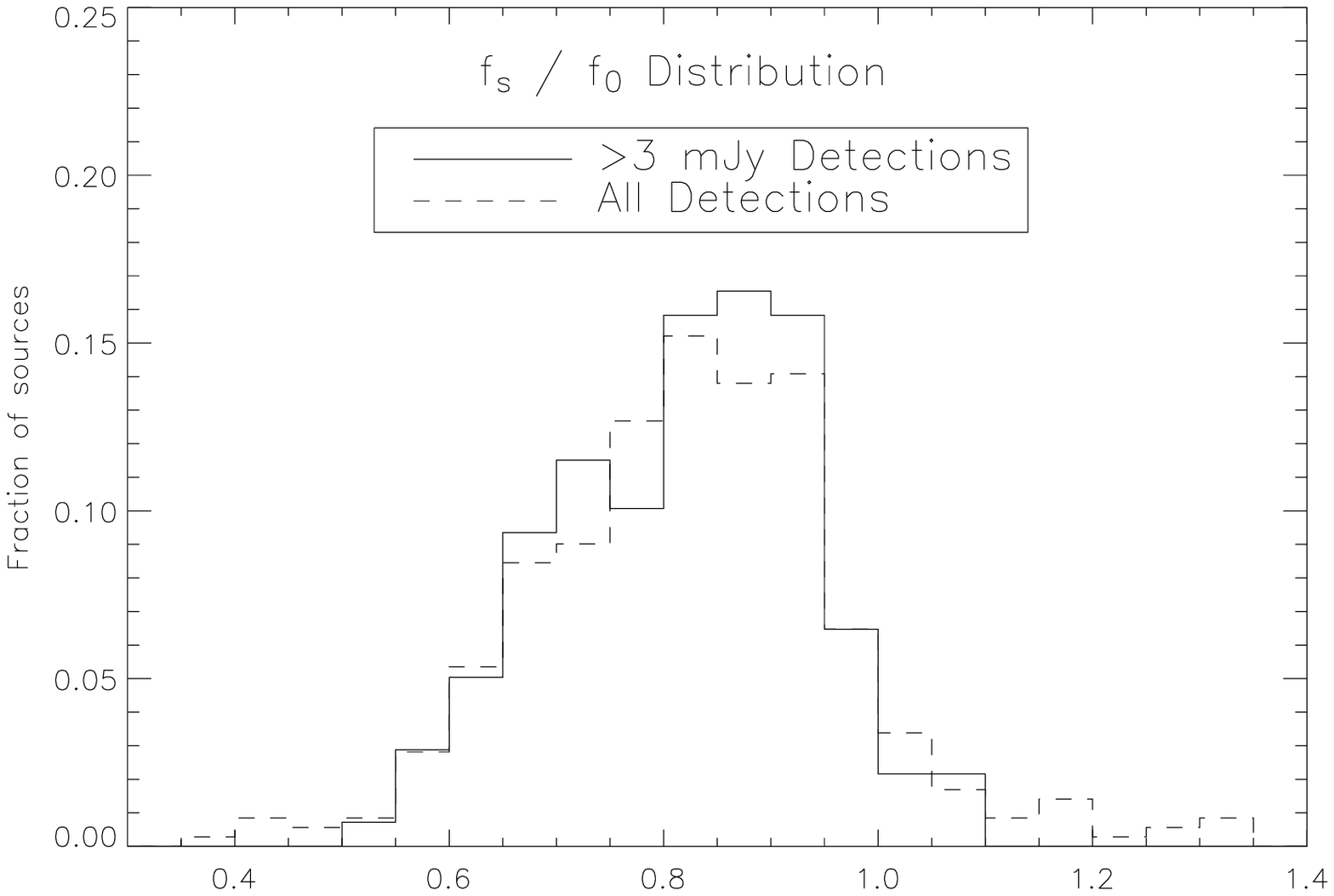}}
\resizebox{0.95\columnwidth}{!}{\includegraphics*{\figdir/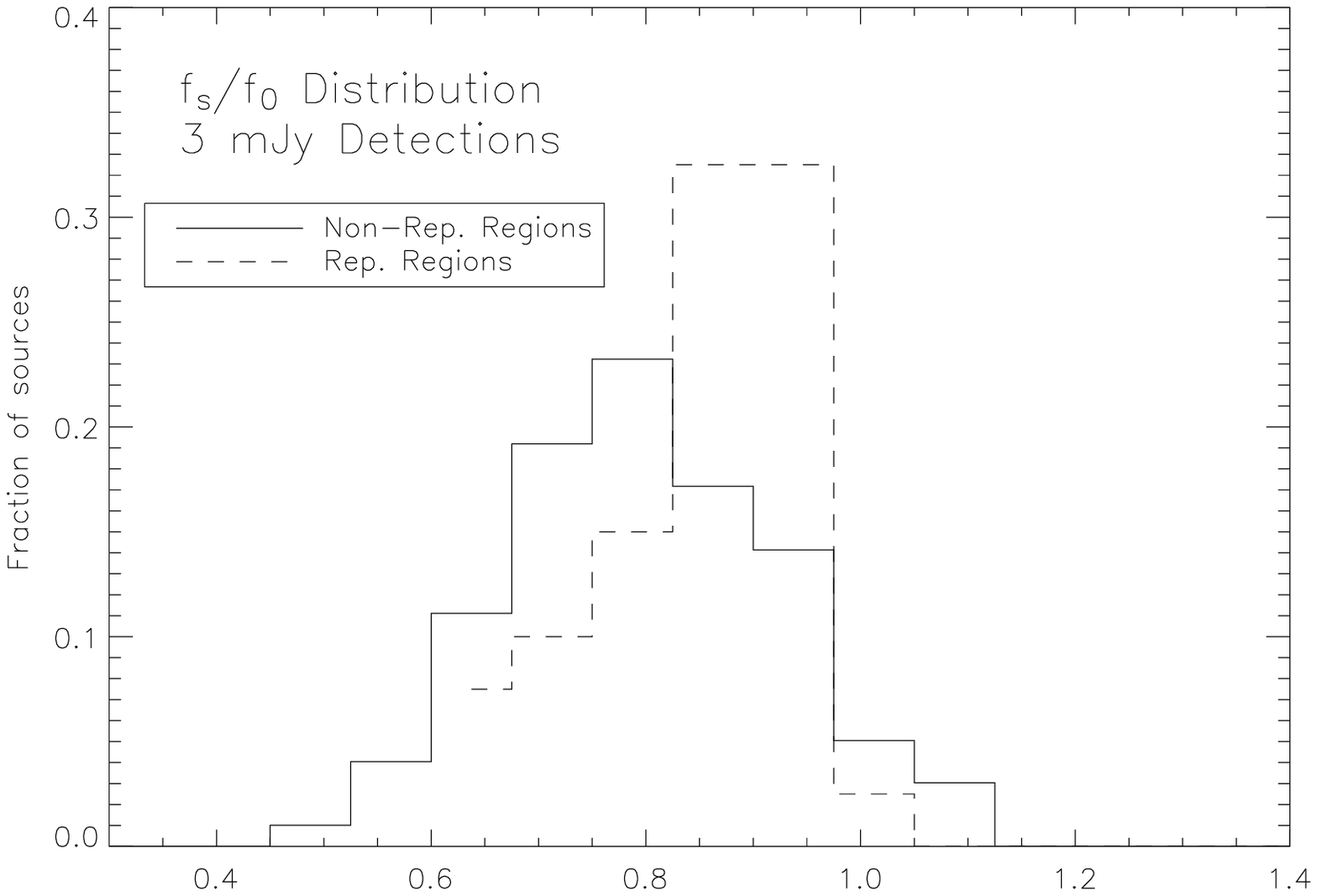}}
\caption{Real/Theoretical Peak Flux Ratio Distribution.
The comparison between distributions for all detections and for bright source
($>$ 3 mJy) detections is shown (top panel) together with the same comparison
for detections in non-repeated and repeated regions (bottom panel).}
\label{fsoverf0.fig}
\end{center}
\end{figure}

\begin{figure}
\begin{center}
\centering
\resizebox{0.95\columnwidth}{!}{\includegraphics*{\figdir/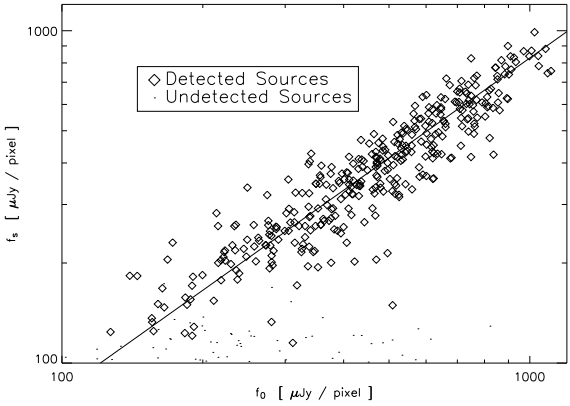}}
\caption{Real vs. Theoretical Peak Flux. The $f_s = 0.83~f_0$ line,
corresponding to the 0.83 mean value of the $f_s/f_0$ ratio for detected
sources, is also plotted.}
\label{fsvsf0.fig}
\end{center}
\end{figure}

In order to correctly determine fluxes, however, one needs to make use
of the employed detector's modeling and thus compare real fluxes with
theoretical ones.
By means of the $f_s/f_0$ histogram distribution, Figure~\ref{fsoverf0.fig}
shows how real peak fluxes are systematically lower than theoretical ones,
due to limits of the data reduction method and to the presence of noise.
The width of the $f_s/f_0$ distribution is quite large, peaking at 0.83,
showing a predictable narrowing of the distribution at bright fluxes
but on the whole a negligible dependence on flux
(Figure~\ref{fsoverf0.fig}, top panel, see also Figure~\ref{fsvsf0.fig}).
Conversely, values for non-repeated and repeated regions differ in a
measurable way, mean values being 0.81 and 0.87, respectively.
(Figure~\ref{fsoverf0.fig}, bottom panel). The rms half-widths of the
distributions are 0.17 and 0.12 in the two cases, respectively, which
also provides an estimate of the photometric error in different regions,
a quantity which will be assessed by different means in
Section~\ref{photoacc.sec}.

Figure~\ref{fsvsf0.fig} confirms the moderately small spread of the
$f_s/f_0$ distribution and also shows the good linearity displayed by
the $f_s$ vs. $f_0$ relation, as long as a source is detected. Provided
a suitable flux correction is applied to account for the $f_s/f_0$
distribution (see Section~\ref{fluxcorr.sec}, this allows to confidently
use our results at all the catalogue flux levels.

\begin{figure}
\begin{center}
\centering
\resizebox{0.95\columnwidth}{!}{\includegraphics*{\figdir/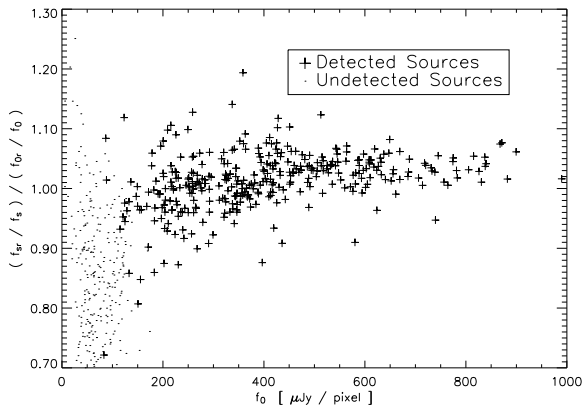}}
\caption{Signal reconstruction as function of real peak flux.}
\label{longratiovsfs.fig}
\end{center}
\end{figure}

Figure~\ref{longratiovsfs.fig} shows how well the signal reconstruction
process carried out by the fitting procedure works at different fluxes.
At it can be clearly seen, below a certain flux, $f_s/f_{sr}$ systematically
falls below the $f_0/f_{0r}$, that is goodness of signal reconstruction
breaks down at the faint end. It was thus decided to use only unreconstructed
fluxes in flux determination, so as to provide a flux estimate that would be
reliable at all flux levels.

\begin{figure}
\begin{center}
\centering
\resizebox{0.95\columnwidth}{!}{\includegraphics*{\figdir/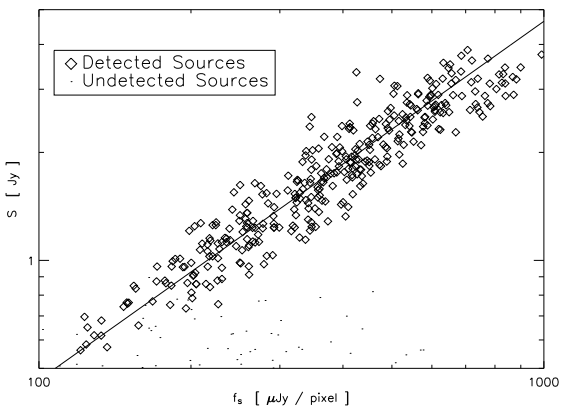}}
\caption{Total vs. Peak Flux. The $S = 4.64~f_s$ line, corresponding to the
4.64 mean value of the $S/f_s$ ratio for detected sources, is also plotted.}
\label{flussivsfs.fig}
\end{center}
\end{figure}

Finally, and most importantly, Figure~\ref{flussivsfs.fig} shows how
total injected fluxes and peak real fluxes are related to each other.
The relation mimics the one between $f_s$ and $f_0$ already shown in
Figure~\ref{fsvsf0.fig}, a linear relation with a reasonably small
spread (the mean value of the $S / f_s$ ratio being 4.64), warranting
a reliable determination of total fluxes on the basis of measured peak
fluxes.
%
\subsection{Flux Correction}~\label{fluxcorr.sec}
\begin{figure}
\begin{center}
\centering
\resizebox{0.95\columnwidth}{!}{\includegraphics*{\figdir/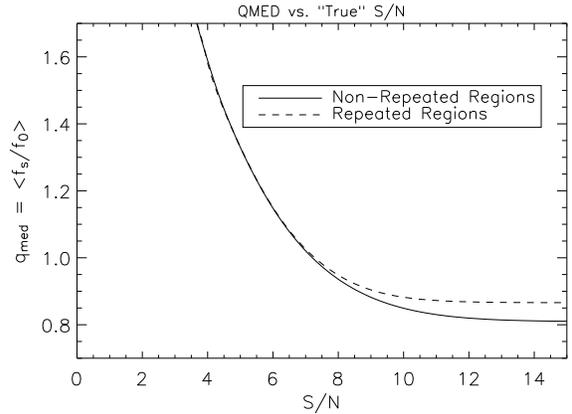}}
\caption{Flux Correction applied to Autosimulated Fluxes. The inverse of the correction factor, or $q_{med}$, is plotted against $S/N$ for non-repeated and repeated regions.}
\label{qmed.fig}
\end{center}
\end{figure}

Systematic effects on flux estimates described in previous Section were
modeled and taken into account using results from simulations and following
\citet{Gruppioni_et_al_2002}.
According to this approach, the correction to be applied to flux estimates
obtained through autosimulation is derived from the the so called $g$ function,
which describes the $S/N$-dependent distribution of the $f_s/f_0$ ratio.
The $g$ function is obtained as the combination of the intrinsic
(i.e. high-$S/N$) $g$ function, or $g_0$ function, and a term due to noise.
First, the $g_0$ function is derived from the distribution of the $f_s/f_0$
ratio obtained for bright ($S > 3~\mathrm{mJy}$) simulated sources (see
Figure~\ref{fsoverf0.fig}), by modeling and correcting for the small degree
of incompleteness to be expected at such fluxes. Then convolution with
a variable noise term is carried out to obtain the overall $S/N$-dependent
$g$ function. Finally, for each $S/N$ value, the median $f_s/f_0$ ratio for
detectable (i.e. $f_s > 5 \sigma$) sources, or $q_{med}$, is computed.
Autosimulated fluxes are then corrected by a factor $1 / q_{med}$.
This process was carried out separately for non-repeated and repeated
regions, and the results are plotted in Figure~\ref{qmed.fig}.
The two $q_{med}$ curves asymptotically tend to the average $f_s/f_0$ values
0.81 and 0.87 determined for simulated sources in non-repeated and repeated
regions, then virtually coincide and soar below $S/N \sim 7$, where
$q_{med} \sim 1$.


It must be emphasized how all of the above applies to "relative" fluxes
determined with respect to ISO LW3 standard sensitivity.
In Section~\ref{photocal.sec} the absolute flux calibration of our catalogue
will be discussed on the basis of the comparison between measured and
predicted MIR stellar fluxes and with respect to IRAS standard calibration.
\subsection{Autosimulation vs. Aperture Photometry}\label{autoaper.sec}
The performance of the autosimulation technique for flux determination
described in Section~\ref{autosim.sec} can also be assessed comparing
the total fluxes obtained with this procedure with those obtained
by means of ordinary aperture photometry.

\begin{figure}
\begin{center}
\centering
\resizebox{0.95\columnwidth}{!}{\includegraphics*{\figdir/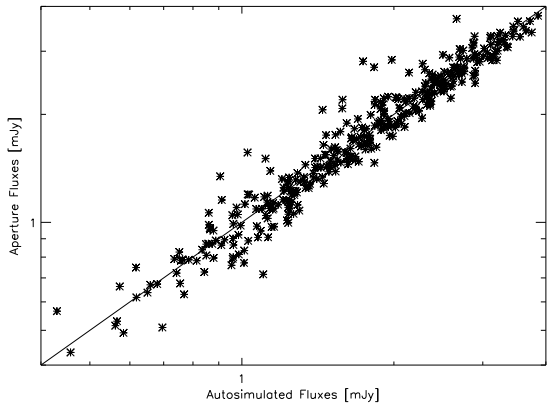}}
\caption{Autosimulated Fluxes vs. Aperture Fluxes}
\label{autoaper.fig}
\end{center}
\end{figure}

The relation between these two quantities shows a good linearity
for most sources, as it is shown for simulated sources in
Figure~\ref{autoaper.fig}.
Plotted aperture fluxes were computed adopting an aperture of 6 arcsec
radius and correcting both for the transients and for the 40\,\% of the
instrumental PSF falling beyond this aperture. Such an aperture was
chosen as the most reliable trade-off allowing to reliably include most
of the source flux and lest of the background.

Only a few sources depart substantially from the 1:1 relation, with 94\,\%
of sources with fluxes in accordance within 20\,\% and an overall rms deviation
of 12\,\%. On the whole, our flux determination procedure is therefore
consistent with conventional aperture photometry, provided a proper
correction for PSF effects is applied.
\subsection{Catalogue Flux Distribution}\label{fluxdist.sec}
\begin{figure}
\begin{center}
\centering
\resizebox{0.95\columnwidth}{!}{\includegraphics*{\figdir/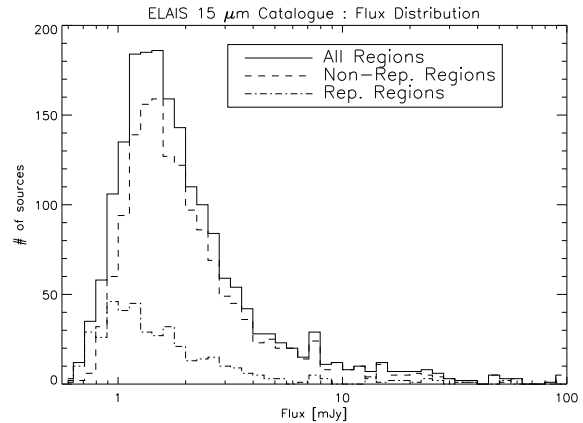}}
\caption{Histogram Flux Distribution of ELAIS Sources. All regions, non-repeated and repeated regions are plotted.}
\label{flux_dist.fig}
\end{center}
\end{figure}

Flux distribution of catalogue sources is illustrated in
Figure~\ref{flux_dist.fig}, with 50\,\% of the sources having fluxes
greater than 1.65 mJy, 76\,\% greater than 1.2 mJy and 89\,\% greater
than 1 mJy. As it can be clearly seen from the two separately plotted
histograms, sources in non-repeated regions largely prevail in number
down to about 1.5 mJy, at which flux their number per flux bin drops
sharply. Conversely, the number of sources per flux bin in repeated
regions continue to increase, if slowly, down to 1 mJy, where they
already amount to about 50\,\% of the total number of sources.
While detailed completeness estimates will be described by
\citet{Lari_et_al_2004}, this plot seems to indicate that the catalogue
is essentially complete up to $\sim$ 1.5 mJy, and slightly fainter than
that in repeated regions.
\section{Optical Identifications}\label{optids.sec}
%
Identification of 15 $\mu$m sources was carried out on both archival and
deep imaging obtained for this purpose at optical and near-infrared
wavelengths.
This provided optimal stellar discrimination as well as optical and
near-infrared flux measurements over the whole large flux range probed
by our catalogue, thus ultimately leading to a very reliable 15 $\mu$m
photometric calibration (see Section~\ref{photocal.sec}).

The S1 field was surveyed in $R$ down to $R \sim 23.0$ by
\citet{LaFranca_et_al_2004} using the 1.5 Danish/ESO telescope.
The N1 and N2 fields were surveyed in $U$, $g'$, $r'$, $i'$ and $Z$
down to 23.4, 24.9, 24.0, 23.2 and 21.9 respectively, as part of the
Wide Field Survey \citep{McMahon_et_al_2001}, using the Wide Field Camera
at the Isaac Newton Telescope. Identifications of N1 and N2 15 $\mu$m sources
based on maximum likelihood were obtained by
\citep{Gonzalez-Solares_et_al_2004}.
Deep near-infrared imaging of areas around selected 15 $\mu$m sources
was carried out by \citep{Vaisanen_et_al_2002}.
Furtherly, USNO A2.0, GSC 2.2, Tycho-2, 2MASS All Sky Data Release and
APM catalogues were cross-correlated with 15 $\mu$m sources so as to provide
a list of tentative identifications.
15 $\mu$m contours were superimposed on DSS and 2MASS images as well as on
new observational material, yielding diagnostic finding charts
for extended, blended and disturbed sources.

Generally speaking, the identification based on deep optical imaging and
automated source extraction and classification carried out with SExtractor was
usually chosen, but archival material proved essential in dealing with bright
sources, which appeared as saturated in deep observations, and in providing
measurements for sources in the N3 or in small portions of the other three
fields, where deep optical imaging had not been obtained. In particular,
a positional difference of 6 arcsec, corresponding to 3 times the maximum
estimated astrometric error of the 15 $\mu$m data reduction process
(see Section~\ref{optids.sec}), was adopted as a cut-off ensuring a safe
identification.

\begin{figure}
\begin{center}
\centering
\resizebox{0.95\columnwidth}{!}{\includegraphics*{\figdir/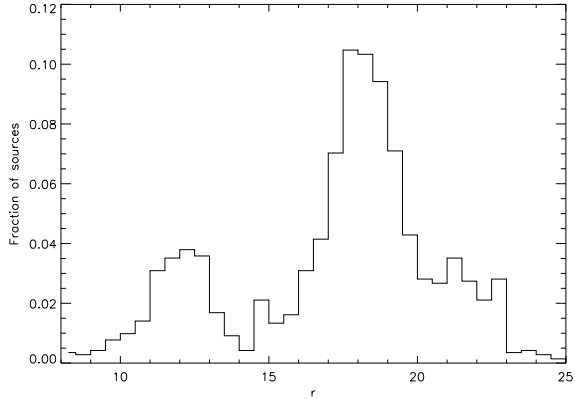}}
\caption{Magnitude distribution of optical IDs of 15 $\mu$m sources. The two peaks
at $r\sim12$ and $r\sim18$ are associated with Stellar and Extragalactic IDs,
respectively.}
\label{rmag_dist_opt_ids.fig}
\end{center}
\end{figure}

The magnitude distribution of the sources identified as optical counterparts
of 15 $\mu$m sources through this process is plotted in histogram form in
Figure~\ref{rmag_dist_opt_ids.fig}, showing a bimodal distribution whose
two peaks are to be associated with stars and galaxies and sharply cutting-off
at $R \sim 23$.
The number of stars and total sources identified according to this criterion,
is separately reported for the four fields in Table~\ref{catalogue2.tab}.
About 90\,\% of 15 $\mu$m sources are assigned a robust optical association
in N1 and N2 fields, while due to the shallower depth this fraction drops
to about 80\,\% in the S1 and N3 fields The average fraction of identified
sources amount to 84.3\,\%. Conversely, the statistics of stellar
identifications is remarkably uniform between the 4 fields, with a 22\,\%
of stars and small field-to-field differences.

\begin{table}
\centering
\caption{Some basic properties of the catalogue divided into different fields.}
\label{catalogue2.tab}
\begin{tabular}{ccccccc}
\hline
Field & Area & Sources & \multicolumn{2}{c}{Stars} & \multicolumn{2}{c}{IDs}\\
      & [deg$^2$] & \# & \# & \% & \# & \% \\
\hline
S1    &  4.17 &  736 & 145 & 19.7 &  584 & 79.3 \\
N1    &  2.84 &  490 & 121 & 24.7 &  441 & 90.0 \\
N2    &  2.84 &  566 & 126 & 22.3 &  493 & 87.1 \\
N3    &  1.00 &  131 &  29 & 22.1 &  103 & 78.6 \\
\hline                                     
Total & 10.85 & 1923 & 421 & 21.9 & 1621 & 84.3 \\
\hline
\end{tabular}
\end{table}

\begin{figure}
\begin{center}
\centering
\resizebox{0.95\columnwidth}{!}{\includegraphics*{\figdir/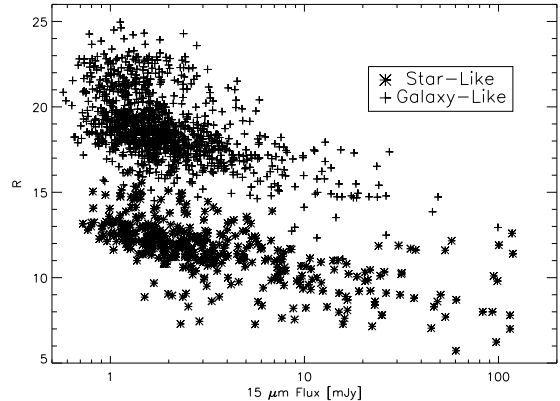}}
\caption{15 $\mu$m vs. $R$ Flux-Flux Diagram of Identified Sources.
Stellar and Extragalactic IDs are indicated.}
\label{1.fig}
\end{center}
\end{figure}

\begin{figure}
\begin{center}
\centering
\resizebox{0.95\columnwidth}{!}{\includegraphics*{\figdir/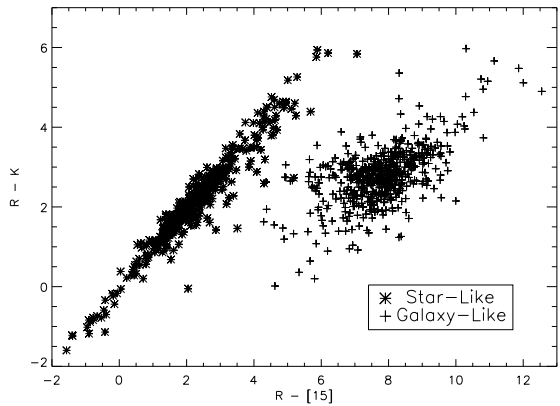}}
\caption{$R-K$ vs. $R-[15]$ Color-Color Diagram of Identified Sources.
Stellar and Extragalactic IDs are indicated.
The $[15]$ magnitude scale is defined in Equation~\ref{15mag.eq}.}
\label{2.fig}
\end{center}
\end{figure}

Figures~\ref{1.fig} and \ref{2.fig} show how, at least down to the 15 $\mu$m
flux limit probed by ELAIS, it is furtherly possible to effectively
discriminate between stars and galaxies solely on the basis of the comparison
between optical and near-infrared magnitudes with 15 $\mu$m fluxes, thus
demonstrating the robustness of the identification process.

The optical identification process also allowed us to assess the astrometric
accuracy achieved in 15 $\mu$m data reduction independently of simulations,
as it is described in Section~\ref{astroacc.sec}.
\section{Astrometric Accuracy}\label{astroacc.sec}
Long-wavelength observations are frequently hampered by large astrometric
errors, leading to severe difficulties in multi-wavelength identifications
and thus physical studies of detected sources.
In our case, astrometric errors can be considered as the combination of
three error sources associated with the detector spatial sampling $\sigma_s$,
the reduction method  $\sigma_r$ and the instrumental pointing accuracy
$\sigma_p$, respectively.

\begin{figure}
\begin{center}
\centering
\resizebox{0.95\columnwidth}{!}{\includegraphics*{\figdir/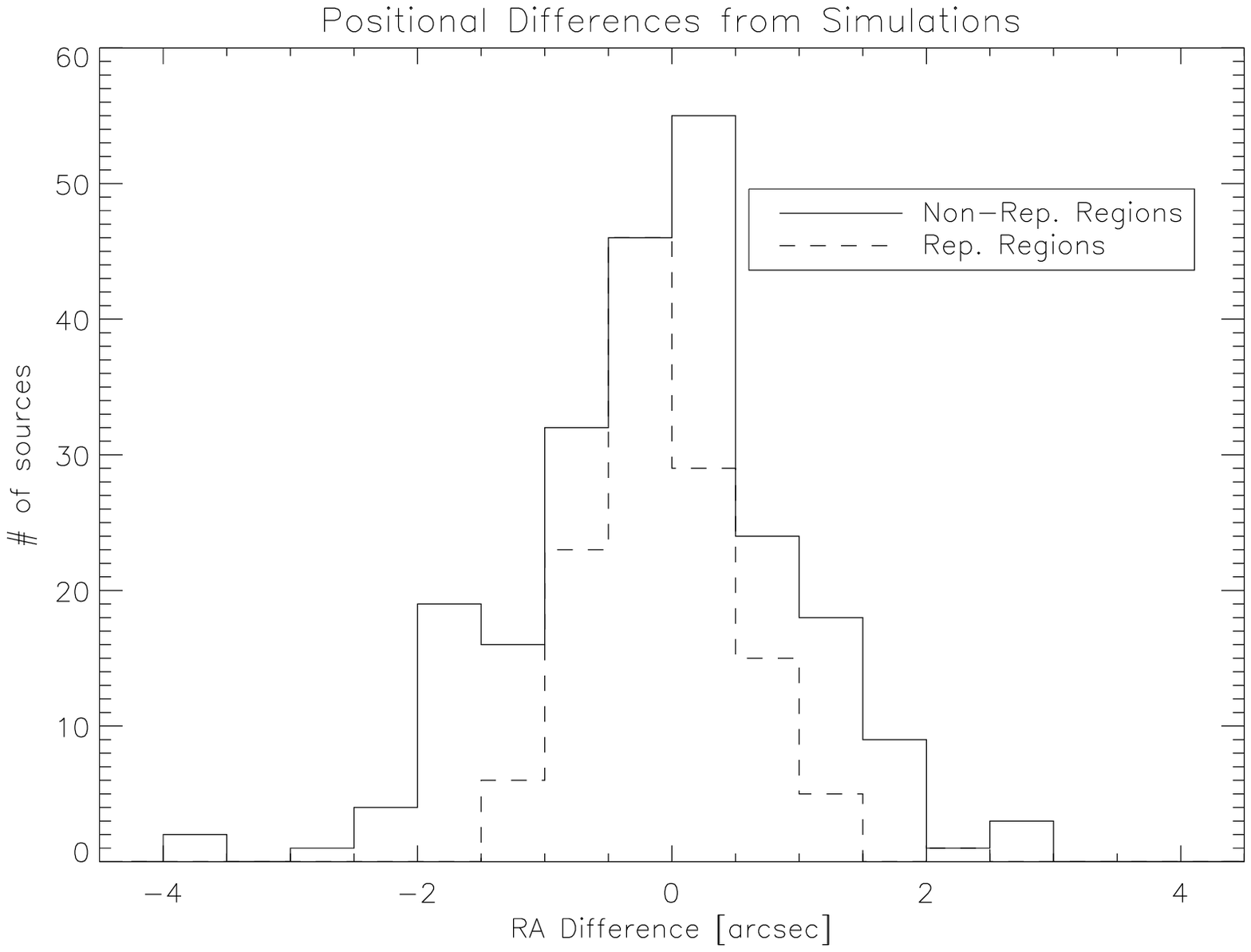}}
\resizebox{0.95\columnwidth}{!}{\includegraphics*{\figdir/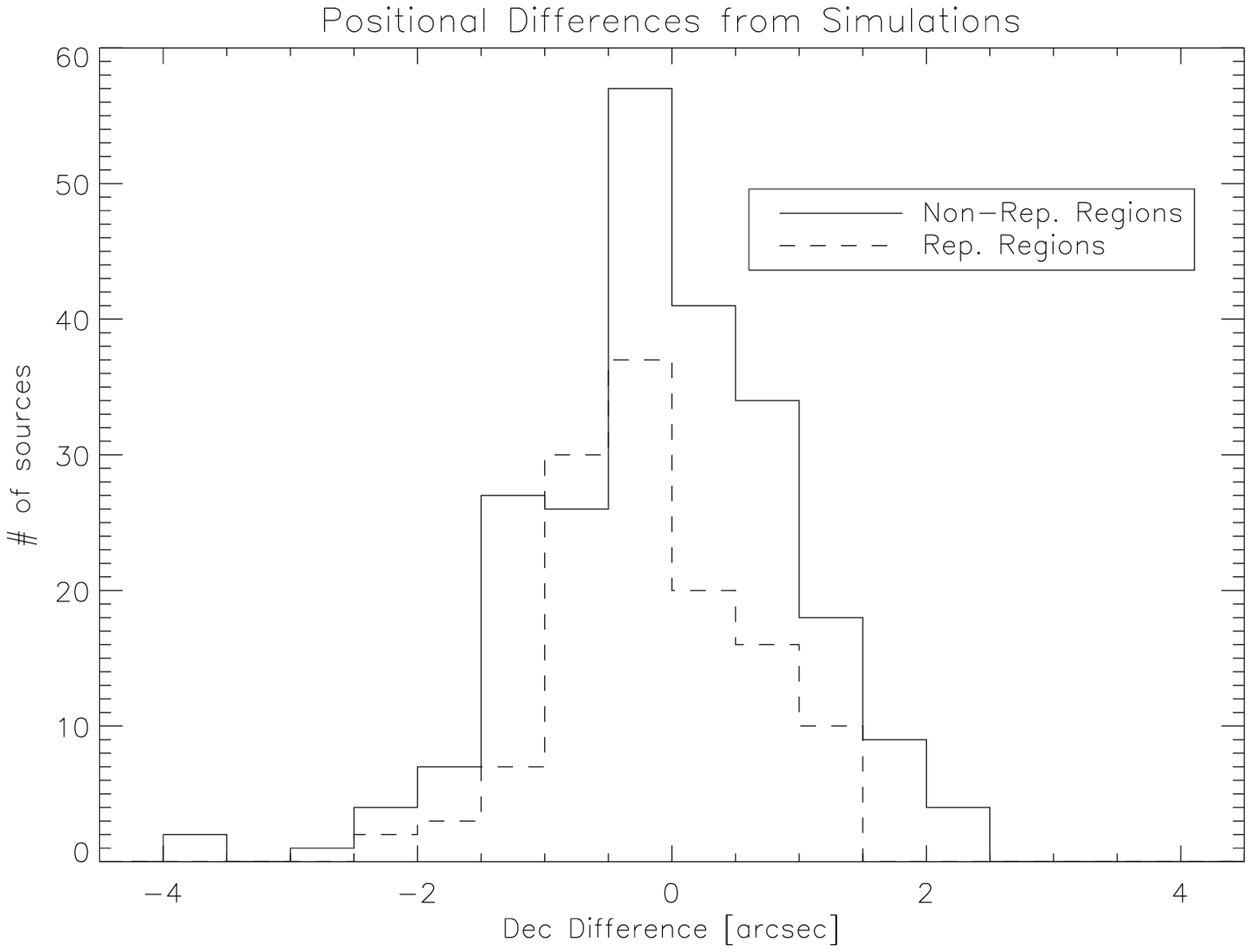}}
\caption{Sampling and Reduction Astrometric Error. Positional differences between injected and detected simulated sources in RA and Dec.}
\label{pos_diff_histo_sim.fig}
\end{center}
\end{figure}

\begin{figure}
\begin{center}
\centering
\resizebox{0.95\columnwidth}{!}{\includegraphics*{\figdir/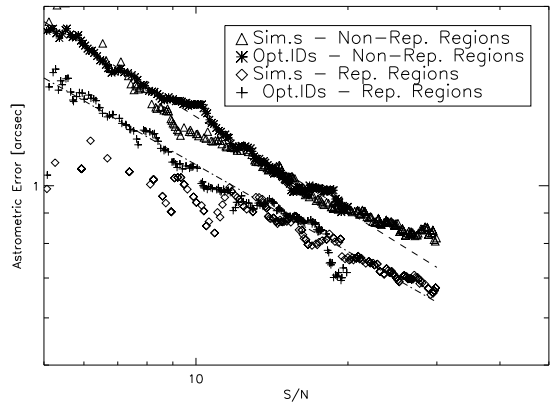}}
\caption{Astrometric Accuracy. Total astrometric errors per component as function of $S/N$ in non-repeated and repeated regions. Polynomial fitting curves used for the evaluation of catalogue astrometric error are also indicated.}
\label{astr_err_sim_and_opt.fig}
\end{center}
\end{figure}

The combination of the $\sigma_s$ and $\sigma_r$ terms, which we will
hereafter indicate as $\sigma_{s+r}$, can be evaluated from simulations,
comparing the injected positions of simulated sources with the corresponding
detected positions, whereas the $\sigma_p$ term is given by the error on
the rasters' astrometric offset, as derived in Section~\ref{mosaicing.sec}.
The total astrometric error will then be given by
\begin{equation}
\sigma_{tot}=\sqrt{\sigma_{s+r}^2+\sigma_p^2}
\label{astr_err.eq}
\end{equation}
The overall distribution of the differences between injected and detected
positions of simulated sources is shown in histogram form in
Figure~\ref{pos_diff_histo_sim.fig} for both non-repeated and repeated regions.
Differences in RA and Dec are distributed in a remarkably similar way, which
will allow us to use them together in the following statistical analysis, so
as to increase our sample. About 75\,\% simulated sources are detected within
1 arcsec of their injected positions and about 96\,\% within 2 arcsec.
Dependence of the $\sigma_{s+r}$ on $S/N$ can be evaluated binning
simulated sources according to their $S/N$ and computing the median
positional difference of each bin. Thus one obtains the $\sigma_{s+r}$
term as function of $S/N$. The $\sigma_p$ term is obtained as the
median value of astrometric offset errors per component reported in
Table~\ref{offsets.tab}, that is 0.39 arcsec per component
irrespectively of $S/N$.
The total astrometric error per component is then computed through
Equation~\ref{astr_err.eq}.
The results are illustrated in Figure~\ref{astr_err_sim_and_opt.fig},
showing total astrometric errors for sources detected both in non-repeated
and repeated regions as function of $S/N$ for $S/N < 30$. The astrometric
accuracy turns out to be very good, errors being as low as 0.7 and 0.8
arcsec per component at the bright end
in non-repeated and repeated regions, respectively.
Catalogue astrometric errors can be estimated by fitting polynomials to the
$\log~\sigma_{tot}$ vs. $\log~S/N$
curves for non-repeated and repeated regions derived from simulations.
At the faint end, however, the errors computed in repeated regions
unrealistically flatten below $S/N \sim 10$, due to the poor statistics
following from the low number of simulated sources that are detected at
low $S/N$ levels.
For this reason, we cross-checked the astrometric accuracy evaluated from
simulations against an independent estimate based on optical identifications
of catalogue sources (see Section~\ref{optids.sec}).
The same binning process described above was carried out on positional
differences between catalogue sources and their optical counterparts,
yielding the results plotted in Figure~\ref{astr_err_sim_and_opt.fig}.
The thus-derived $\sigma_{tot}$ vs. $S/N$ relation closely follows a straight
line in the log-log plane at $5 < S/N < 20$, and closely resembles the one
derived from simulations at $10 < S/N < 20$, confirming the overall
robustness of the simulation process.
It was thus decided to estimate catalogue astrometric errors by fitting a
straight line to the $\log~\sigma_{tot}$ vs. $\log~S/N$ curves for
non-repeated and repeated regions derived from optical identifications.
At $S/N > 20$, other effects such as the widening of the instrumental PSF
and the increase of the optical astrometric error
begin dominating the astrometric error budget, and the astrometric
error curve starts showing irregularities, so that linear fitting
cannot be reliably assumed to describe the actual astrometric error.
Accordingly, values obtained from simple extrapolation of the straight line
obtained at $S/N < 20$ to higher $S/N$ were truncated when they were deemed
too optimistic, i.e. at the 0.8 and 0.7 arcsec values for non-repeated and
repeated regions, corresponding to the extrapolated values for $S/N \sim 25$.
%
%
%
\section{Photometric Accuracy}\label{photoacc.sec}
\begin{figure}
\begin{center}
\centering
\resizebox{0.95\columnwidth}{!}{\includegraphics*{\figdir/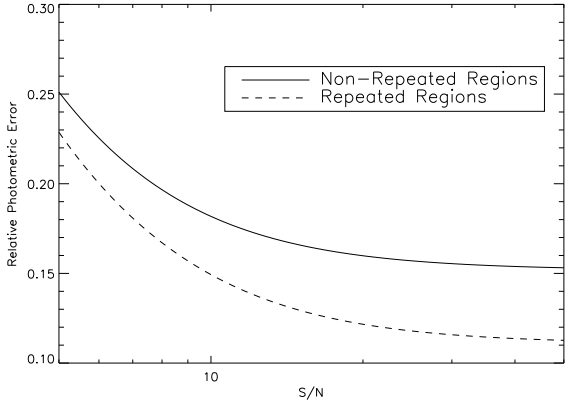}}
\caption{Relative photometric error as function of $S/N$ as given by
Equation~\ref{relphoterr.eq} for non-repeated and repeated regions.}
\label{relphoterr.fig}
\end{center}
\end{figure}

Errors in flux determination using our method can mainly be attributed
to two effects, namely the autosimulation process and the noise present
on the sky maps.
The first contribution can be estimated by computing the width of the
$f_s/f_0$ distribution shown in Figure~\ref{fsoverf0.fig} for high $S/N$
sources only, so as to evaluate the effects of the autosimulation process
on relatively noise-free maps. At lower $S/N$, photometric errors arising
from noise on the sky maps will combine with those arising from autosimulation.
The overall error is thus given by
\begin{equation}
\left(\frac{\Delta S}{S}\right)^2 = \Delta\left(\frac{f_s}{f_0}\right)^2 + \left(\frac{\sigma_{map}}{f_s}\right)^2 = \Delta\left(\frac{f_s}{f_0}\right)^2 + \left(\frac{1}{S/N}\right)^2
\label{relphoterr.eq}
\end{equation}
where $\Delta(f_s/f_0)$ is the width of the $f_s/f_0$ distribution
as measured for high $S/N$ sources for non-repeated regions and
repeated regions separately, and $\sigma_{map}$ is the noise
as measured at each source position on the sky map. The first term
is a constant, about 0.15 and 0.11 for non-repeated and repeated
regions, respectively. This term dominates the photometric error budget
at high $S/N$, whereas the importance of the second term increases when
$S/N$ decreases.
Photometric errors estimated using Equation~\ref{relphoterr.eq} are shown
in Figure~\ref{relphoterr.fig}.
The values for $S/N \sim 10$ are 0.18 / 0.15 for non-repeated / repeated
regions, while they increase to 0.25 / 0.23 for $S/N \sim 5$.
\section{Photometric Calibration}~\label{photocal.sec}
The standard sensitivity of ISO LW3 channel is of 1.96 ADU/Gain/s/mJy,
which can be used to calibrate our catalogue fluxes. However, given the
large sample at our disposal, the accuracy of such photometric
calibration can be tested against the independently determined IRAS standard
photometric calibration by studying detected sources with stellar counterparts.
\citet{Aussel_2004} performed a detailed study of mid-infrared emission
from stars, using large samples drawn from IRAS Faint Source Catalog with
counterparts in the 2MASS and Tycho-2 \citep{Hoeg_et_al_2000}
catalogues.
In the two cases, it is found that the $J-K$ and $B-V$ colors of stars
are extremely well correlated with the $K-[12]$ and $B-[12]$ colors,
respectively, where $[12]$ is a magnitude scale constructed from
IRAS 12 $\mu$m flux, following the prescriptions of \citet{Omont_et_al_1999}.
This relation allows to accurately predict the IRAS 12 $\mu$m flux of a star,
provided that its $J$ and $K$ (or $B$ and $V$) are known and that
$J-K$ (or $B-V$) are within certain limits.
Stellar atmosphere models \citep{Lejeune_et_al_1998} show that such color
criteria select stellar spectral types for which the ratio between the
15 $\mu$m flux and the 12 $\mu$m flux is essentially constant.
Thus, if one has access to a substantial number of optical and/or near-infrared
counterparts to 15 $\mu$m sources, it is possible to use the correlations
by \citet{Aussel_2004} to predict the 15 $\mu$m fluxes of ISO-detected stars
and compare them to the measured values. Such a comparison then allows to
investigate possible systematic differences between the two sets of fluxes,
which should be attributable to discrepancies between the independently
established IRAS and ISO calibrations.
Given the tighter nature of the near-infrared/mid-infrared correlation
over a wide range of fluxes with respect to the optical/mid-infrared one,
it was decided to use the former, that is expressed by
\begin{equation}
K - [15] = 0.044 + 0.098~( J - K ) ~,
\end{equation}
where $[15]$ is a magnitude scale defined as
\begin{equation}
[15] = 3.202 - 2.5~\log~( S_{15~\mu\mathrm{m}}~[\textrm{mJy}]~)~,
\label{15mag.eq}
\end{equation}
to compute predicted fluxes.
2MASS All Sky Release fluxes were used for the majority of the sources
together with fluxes determined for fainter sources as part of the
near-infrared follow-up program to ELAIS observations
\citep{Vaisanen_et_al_2002}.
\begin{figure}
\begin{center}
\centering
\resizebox{0.95\columnwidth}{!}{\includegraphics*{\figdir/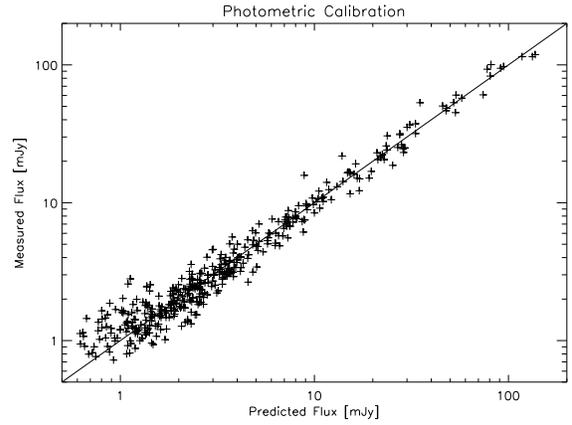}}
\caption{Photometric calibration. Measured vs. Predicted 15 $\mu$m Stellar
Fluxes. The 1.0974 correction factor to measured fluxes is here already taken
into account, and the one-to-one relation is also plotted.}
\label{phot_cal_stars.fig}
\end{center}
\end{figure}
In Figure~\ref{phot_cal_stars.fig} measured fluxes are plotted against
predicted ones for all 408 sources with reliable stellar identification
and near-infrared magnitudes, showing the remarkable linearity and the
small spread of the relation over a wide range of fluxes.
\begin{figure}
\begin{center}
\centering
\resizebox{0.95\columnwidth}{!}{\includegraphics*{\figdir/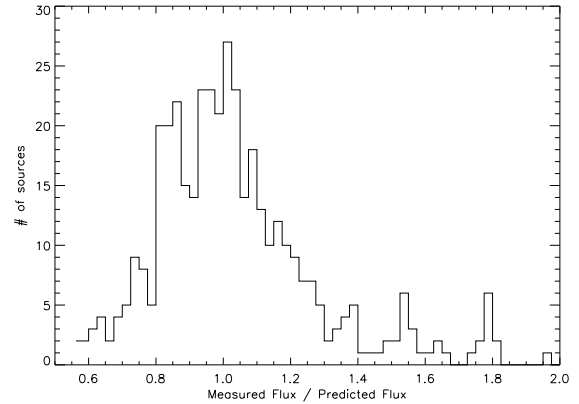}}
\caption{Photometric calibration. Histogram Distribution of Measured / Predicted
15 $\mu$m Stellar Flux Ratio. The 1.0974 correction factor to measured fluxes
is here already taken into account.}
\label{phot_cal_stars_histo.fig}
\end{center}
\end{figure}
Figure~\ref{phot_cal_stars_histo.fig} shows the histogram distribution of the
measured/predicted flux ratio.
Reducing the sample to the 300 sources with a $S/N > 10$ yields an average
predicted/measured ratio of 1.0974 with a standard deviation of 0.0121.
The difference between measured and predicted fluxes is significant at $\sim 8$
sigma, suggesting the presence of systematic effects in either the IRAS or ISO
calibration process.
Disentangling the effects leading to such a discrepancy would involve
the reduction of a wide set of IRAS-detected sources observed by ISO using
our method.
Given the substantial amount of work that would be needed for this purpose,
it was decided to simply correct our catalogue fluxes by a constant 1.0974
factor to put them on the IRAS scale.
However, while this choice was taken for the sake of compatibility of our
results with studies adopting the IRAS flux scale (and particularly
IRAS-based source counts and luminosity functions), this is
not to indicate that IRAS calibration
is more secure that ISO's.
In fact, results from data reduction of deep ISOCAM surveys in the
Lockman Hole using the LARI Method \citep{Fadda_et_al_2004} favour
the ISO calibration.
On the basis of stellar atmosphere models applied to a smaller number of
stars and deep multi-band imaging,
the authors find that model predictions agree with 15 $\mu$m fluxes
determined following ISO calibration. 
%
%
\section{The Catalogue}\label{catalogue.sec}
The ELAIS 15 $\mu$m Final Analysis Catalogue (Version 1.0) contains 1923
sources detected with a $S/N$ greater than 5 in the (RA-ordered) four fields
S1, N3, N1 and N2, totalling an area of $10.18~\mathrm{deg}^2$.
For each entry, the catalogue reports astrometric and photometric information,
optical identification and a number of ancillary flags.
The entries are as detailed in the following:
\begin{itemize}
\item{\textbf{Name}} : IAU source name constructed from Right Ascension and Declination;
\item{\textbf{RA (J2000)}} : Right Ascension at epoch J2000 in both decimal and sexagesimal angular units;
\item{\textbf{Dec (J2000)}} : Declination at epoch J2000 in both decimal and sexagesimal angular units;
\item{\textbf{Total Flux}} : source total flux obtained from autosimulation or aperture photometry (see "Aperture" below), expressed in mJy;
\item{\textbf{Peak Flux}} : source peak flux measured on unreconstructed maps, expressed in $\mu$Jy/pixel;
\item{\textbf{S/N}} : signal-to-noise ratio measured on unreconstructed maps
\item{\textbf{Astrometric Error}} : Astrometric error as determined from Equation~\ref{astr_err.eq}
\item{\textbf{Photometric Error}} : Photometric error as determined from Equation~\ref{relphoterr.eq}
\item{\textbf{Aperture}} : aperture photometry flag, 0 for sources whose flux was estimated through autosimulation and 1 for sources on which aperture photometry was preferred (see Section~\ref{autosim.sec});
\item{\textbf{Optical ID}} : optical identification as described in Section~\ref{optids.sec}. S stands for star, G for galaxy and * for unidentified source. A further D flags a particularly dubious (non-)identification.
\end{itemize}
A sample of the catalogue, which will shortly become available at
\texttt{http://astro.imperial.ac.uk/$\sim$vaccari/elais}, is shown in
Table~\ref{catalogue.tab}.
Some of the data on the sources are also contained within the ELAIS
band-merged catalogue by \citet{Rowan-Robinson_et_al_2003}.
\begin{table*}
\centering
\caption{A sample portion of ELAIS 15 $\mu$m Final Analysis Catalogue.
Source IAU Name, Right Ascension and Declination at J2000 epoch in
sexagesimal formats, Total and Peak Flux, Signal-To-Noise Ratio,
Astrometric and Photometric Errors, Aperture Flag.
Note that a few columns described in text were dropped for the sake
of brevity.
}
\label{catalogue.tab}
\begin{small}
\begin{tabular}{ccccccccc}
\hline
Name &  RA (J2000)  &  DEC (J2000)  & Total F.   & Peak F. & $S/N$ & Astr. E. & Phot. E. & Ap. \\
\hline
 & hr~~min~~sec & deg~~min~~sec & mJy & $\mu$Jy & & arcsec & mJy & \\
\hline
ELAISC15\_J143048.92+332830.08 &  14 30 48.9265 &  33 28 30.0837 &   1.2203 &   187.52 &   6.375 &  1.657 &  0.266 &  0 \\
ELAISC15\_J143053.24+333119.52 &  14 30 53.2422 &  33 31 19.5201 &   4.2931 &   649.40 &  18.421 &  0.944 &  0.691 &  0 \\
ELAISC15\_J143059.16+332215.08 &  14 30 59.1625 &  33 22 15.0859 &   3.0696 &   565.34 &  16.526 &  1.000 &  0.501 &  0 \\
ELAISC15\_J143112.10+325625.87 &  14 31 12.1003 &  32 56 25.8750 &   1.6166 &   166.20 &   5.266 &  1.834 &  0.392 &  0 \\
ELAISC15\_J143116.96+331957.83 &  14 31 16.9618 &  33 19 57.8304 &   3.1798 &   657.70 &  27.828 &  0.700 &  0.370 &  0 \\
ELAISC15\_J143116.99+332903.79 &  14 31 16.9954 &  33 29  3.7930 &   1.3766 &   296.39 &  12.327 &  0.980 &  0.189 &  0 \\
ELAISC15\_J143122.40+332036.07 &  14 31 22.4041 &  33 20 36.0725 &   2.2152 &   356.15 &  10.501 &  1.272 &  0.396 &  0 \\
ELAISC15\_J143123.57+330517.23 &  14 31 23.5781 &  33  5 17.2354 &  57.3303 & 11957.87 & 365.708 &  0.800 &  8.693 &  0 \\
ELAISC15\_J143125.36+331348.76 &  14 31 25.3603 &  33 13 48.7608 &  26.3799 &  1795.13 &  53.649 &  0.800 &  4.029 &  1 \\
ELAISC15\_J143131.33+330143.80 &  14 31 31.3313 &  33  1 43.8084 &   6.7570 &   958.24 &  30.352 &  0.800 &  1.048 &  0 \\
ELAISC15\_J143135.38+333224.62 &  14 31 35.3831 &  33 32 24.6235 &   0.8125 &   122.88 &   5.278 &  1.477 &  0.178 &  0 \\
ELAISC15\_J143135.47+325456.62 &  14 31 35.4741 &  32 54 56.6245 &   1.6331 &   301.73 &   9.053 &  1.376 &  0.306 &  0 \\
ELAISC15\_J143137.69+325453.32 &  14 31 37.6931 &  32 54 53.3249 &   2.5355 &   495.93 &  14.764 &  1.062 &  0.421 &  0 \\
ELAISC15\_J143138.44+332808.85 &  14 31 38.4461 &  33 28  8.8581 &   1.4309 &   274.79 &   7.864 &  1.483 &  0.283 &  0 \\
ELAISC15\_J143140.70+330316.82 &  14 31 40.7051 &  33  3 16.8265 &   3.6781 &   747.97 &  23.415 &  0.831 &  0.579 &  0 \\
ELAISC15\_J143142.01+331003.66 &  14 31 42.0114 &  33 10  3.6686 &   1.6108 &   269.70 &   8.391 &  1.433 &  0.310 &  0 \\
ELAISC15\_J143143.08+325301.48 &  14 31 43.0892 &  32 53  1.4801 &   1.2484 &   249.18 &   5.679 &  1.762 &  0.290 &  0 \\
ELAISC15\_J143143.58+333200.05 &  14 31 43.5875 &  33 32  0.0592 &   2.4374 &   386.42 &  14.690 &  1.065 &  0.405 &  0 \\
ELAISC15\_J143143.69+330133.41 &  14 31 43.6981 &  33  1 33.4117 &   2.4854 &   393.08 &  12.373 &  1.166 &  0.427 &  0 \\
ELAISC15\_J143143.86+333119.96 &  14 31 43.8619 &  33 31 19.9622 &   0.9966 &   200.88 &   6.450 &  1.647 &  0.216 &  0 \\
ELAISC15\_J143149.61+330212.86 &  14 31 49.6144 &  33  2 12.8683 &   2.7584 &   404.89 &  14.698 &  1.064 &  0.458 &  0 \\
ELAISC15\_J143155.98+330138.26 &  14 31 55.9881 &  33  1 38.2622 &   1.5529 &   263.79 &   7.163 &  1.558 &  0.320 &  0 \\
ELAISC15\_J143156.34+325138.33 &  14 31 56.3402 &  32 51 38.3395 &   3.8175 &   811.95 &  17.343 &  0.975 &  0.619 &  0 \\
ELAISC15\_J143159.54+325439.10 &  14 31 59.5441 &  32 54 39.1004 &   1.3582 &   232.94 &   6.381 &  1.656 &  0.296 &  0 \\
ELAISC15\_J143201.02+331525.84 &  14 32  1.0279 &  33 15 25.8460 &   2.0073 &   342.84 &   9.346 &  1.353 &  0.372 &  0 \\
\hline
\end{tabular}
\end{small}
\end{table*}
\section{Conclusions}\label{conclusions.sec}
A technique for ISO-CAM/PHOT data reduction, the LARI method, was variously
refined and applied to ELAIS 15 $\mu$m observations.
The mathematical model for the detector's behaviour is the same as
originally presented in \citet{Lari_et_al_2001}, but thanks to various
improvements, and particularly to a new Graphical User Interface, the
method is now more robust and, most importantly, quicker and easier to apply
to large datasets.
Its application, in the new form, to the four fields composing the dataset
(including a re-reduction of S1 observations already presented in
\citet{Lari_et_al_2001}) has produced a catalogue of 1923 sources
spanning the 0.5 -- 100 mJy range, detected with a $S/N$ greater than 5
over a total area of $10.85~\mathrm{deg}^2$.
Optical identification of 15 $\mu$m sources has been carried out on
heterogeneous optical and near-infrared imaging material, allowing to
determine a robust association for about 85\,\% of the sources and
identify 22\,\% of them as bona fide stars, furtherly demonstrating
the reliability of our data reduction process.

The evaluation of the catalogue's quality has been carried out through
both accurate simulations and multi-wavelength identification.
The astrometric accuracy is of order 1 arcsec in both RA and Dec for
$S/N > 10$, while it increases up to about 2 arcsec in both RA and Dec for
$S/N \sim 5$, and somewhat better for sources detected in higher-redundancy
sky regions.
The photometric accuracy is estimated to be below 25\,\% over the whole range
of fluxes and redundancy levels probed by our catalogue, and better than 15\,\%
for $S/N > 10$ sources.

The comparison of measured stellar fluxes with fluxes estimated on the basis
of stellar atmospehere models calibrated on IRAS data and on near-infrared
photometry allowed to achieve an IRAS/ISO relative photometric calibration.
An IRAS/ISO relative calibration factor of $1.0974 \pm 0.0121$ was determined,
shedding doubts on the goodness of the two independently determined
calibrations at the 10\,\% level. For lack of a simple way to identify
error sources in IRAS and/or ISO calibratioon process, it was decided
to put our catalogue on the more commonly used IRAS flux scale.

In a forthcoming paper \citep{Lari_et_al_2004} completeness estimates
and extragalactic source counts from this catalogue will be presented,
covering the crucial flux range 0.5 -- 100 mJy between ISOCAM 15 $\mu$m
Deep Surveys and IRAS All Sky Survey.
%
%
%
\section*{Acknowledgments}
This paper is based on observations with ISO, an ESA project with instruments
funded by ESA Member States (especially the PI countries: France, Germany,
the Netherlands and the United Kingdom) and with the participation of ISAS
and NASA.

The ISOCAM data presented in this paper were analysed using CIA, a joint
development by the ESA Astrophysics Division and the ISOCAM Consortium.
The ISOCAM Consortium is led by the ISOCAM PI, C. Cesarsky.


This work was partly supported by the "POE" EC TMR Network Programme
(HPRN-CT-2000-00138).
%

%
%
%
%
\bsp
\label{lastpage}
\end{document}